\RequirePackage{fix-cm}
\documentclass[smallextended]{svjour3}       
\smartqed  
\usepackage{epsfig}
\usepackage{graphicx}
\usepackage{amsmath, amssymb}
\usepackage{xcolor}

\newcommand{\Real}{\mathbb{R}}

\newcommand{\Volume}{\mathcal{V}}

\definecolor{mycolor}{rgb}{0.7,0.3,0.3}

\newcommand{\bfw}{\boldsymbol{w}}
\newcommand{\bfs}{\boldsymbol{s}}
\newcommand{\bfx}{\boldsymbol{x}}

\journalname{Bull Math Biol}

\begin{document}

\title{High-dimensional brain}
\subtitle{A tool for encoding and rapid learning of memories by single neurons}


\author{Ivan Tyukin         \and
        Alexander N. Gorban \and Carlos~Calvo \and Julia~Makarova \and Valeri~A.~Makarov
}


\institute{I. Tyukin \at
              University of Leicester, Department of Mathematics, University Road, LE1 7RH, United Kingdom \\
              Tel.: +44-116-2525106\\
              \email{I.Tyukin@le.ac.uk}\\
              Saint-Petersburg State Electrotechnical University, Saint-Petersburg, Prof. Popova str. 5, Russia
           \and
           Alexander N. Gorban \at
              University of Leicester, Department of Mathematics, University Road, LE1 7RH, United Kingdom
           \and
           Carlos Calvo \at
           Instituto de Matem\'atica Interdisciplinar, Faculty of Mathematics, Universidad Complutense de Madrid, Avda Complutense s/n, 28040 Madrid, Spain
           \and
           Julia Makarova \at
               Department of Translational Neuroscience, Cajal Institute – CSIC, Madrid, Spain;\\
               Lobachevsky State University of Nizhny Novgorod, Gagarin Ave. 23, 603950 Nizhny Novgorod, Russia
           \and
           Valeri A. Makarov \at
           Instituto de Matem\'atica Interdisciplinar, Faculty of Mathematics, Universidad Complutense de Madrid, Avda Complutense s/n, 28040 Madrid, Spain;\\
           Lobachevsky State University of Nizhny Novgorod, Gagarin Ave. 23, 603950 Nizhny Novgorod, Russia\\
           \email{vmakarov@ucm.es}
}

\date{Received: date / Accepted: date}

\maketitle

\begin{abstract}
Codifying memories is one of the fundamental problems of modern Neuroscience. The functional mechanisms behind {this phenomenon} remain largely unknown. {Experimental} evidence suggests that some of the memory functions are performed by stratified brain structures such as, e.g., the hippocampus. In this particular case, single neurons in the CA1 region receive a highly multidimensional input from the CA3 area, which is a hub for information processing. We thus assess the implication of the abundance of neuronal signalling {routes} converging onto single cells on the information processing. We show that single neurons can selectively detect and learn arbitrary information items, given that they operate in high dimensions. The argument is based on Stochastic Separation Theorems and {the concentration of measure} phenomena. We demonstrate that a simple enough functional neuronal model is capable of explaining: i) the extreme selectivity of single neurons to the information content, ii) simultaneous separation of several uncorrelated stimuli or informational items from a large set, and iii) dynamic learning of new items by associating them with already ``known'' ones. These results constitute a basis for organization of complex memories in ensembles of single neurons. Moreover, they show that no \textit{a priori} assumptions on the structural organization of neuronal ensembles are necessary for explaining basic concepts of static and dynamic memories.

\keywords{Neural memories \and Single-neuron learning \and Perceptron \and Stochastic Separation Theorems}
\end{abstract}

\section{Introduction} \label{intro}

The human brain is arguably amongst the most sophisticated and enigmatic nature creations. Over millions of years it has evolved to amass  billions of neurons, featuring on average $86 \times 10^9$ cells  \cite{Herculano2012}. This remarkable figure is several orders of magnitude higher than that of the most mammals and several times larger than in primates \cite{Herculano2011}. Whilst measuring roughly $2\%$ of the body mass, the human brain consumes about $20\%$ of the total energy \cite{Clark1999}.

The significant metabolic cost associated with a larger brain in humans, as opposed to mere body size - a path that great apes might have evolved  \cite{Herculano2011}, must be justified by evolutionary advantages. Some of the benefits may be related to the development of a remarkably important social life in humans. This, in particular, requires extensive abilities in formation of complex memories. Indirectly this hypothesis is supported by the significant difference among species in the number of neurons in the cortex \cite{Herculano2009} and the hippocampus \cite{Andersen2007}. For example, in the CA1 area of the hippocampus there are $0.39\times 10^6$ pyramidal neurons in rats, $1.3\times 10^6$ in monkeys, and $14\times 10^6$ in humans.

Evolutionary implications in relation to cognitive functions have been widely discussed in the literature (see, e.g., \cite{Platek2007,Sherwood2012,Sousa2017}). Recently, it has been shown  that in humans new memories can be learnt very rapidly by supposedly individual neurons from a limited number of experiences \cite{Quiroga:2015}. Moreover, some neurons can exhibit remarkable selectivity to complex stimuli, the evidence that has led to debates around the existence of the so-called ``grand mother'' and ``concept''  cells \cite{Quiroga:2005,Quiroga:2009,Quiroga:2012}, and their role as elements of a declarative memory. These findings suggest that not only the brain can learn rapidly but also it can respond selectively to ``rare'' individual stimuli. Moreover,  experimental evidence indicates that such a cognitive  functionality can be delivered by  single neurons \cite{Quiroga:2015,Quiroga:2005,Quiroga:2009}. The fundamental questions, hence, are: How is this possible? and What could be the underlying functional mechanisms?

Recent theoretical advances achieved within the Blue Brain Project show that the brain can operate in many dimensions \cite{Markram2017}. It is claimed that the brain has structures operating in up to eleven dimensions. Groups of neurons can form the so called \textit{cliques}, i.e., networks of specially interconnected neurons that generate precise representations of geometric objects. Then the dimension grows with the number of neurons in the clique. {Multidimensional representation of spatiotemporal information in the brain is also implied in the concept of} generalized cognitive maps (see, e.g., \cite{BC2015,ACS2016,LimbMov}). Within this theory, {spatiotemporal relations between objects in the environment are encoded as static (cognitive) maps and represented as elements of an $n$-dimensional space ($n\gg 1$). The cognitive maps as information items can be learnt, classified, and retrieved on demand \cite{IEEE2013}.} However, the questions {concerning} how the brain or individual neurons can distinguish among a huge number of different maps and select an appropriate one remain unknown.

In this work we propose that brain areas with a predominant laminar topology and abundant signalling {routes} simultaneously converging on individual cells (e.g., the hippocampus) are propitious for a high-dimensional processing and learning of complex information items. We show that a canonical neuronal model,  the perceptron \cite{Rosenblatt1962}, in combination with a Hebbian-type of learning  may provide answers to the above mentioned fundamental questions. In particular, starting from stochastic separation theorems \cite{GorbanTyukin:NN:2017,GorbanTyukin:RSTA:2017} we demonstrate that  individual neurons gathering multidimensional stimuli through a sufficiently large number of synaptic inputs can exhibit extreme selectivity either to individual information items or to groups of items. Moreover, neurons are capable of associating and learning uncorrelated information items.  Thus, a large number of signalling {routes} simultaneously converging on a large number of single cells, as it is widely observed in laminar brain structures, translates into a natural environment for rapid formation and maintenance of extensive memories. This is vital for social life, and hence may constitute a significant evolutionary advantage, albeit, at the cost of high metabolic expenditure.

\section{Fundamental problems of encoding memories}\label{sec:problem}

\begin{figure}[!h]
\centering{\epsfig{file = 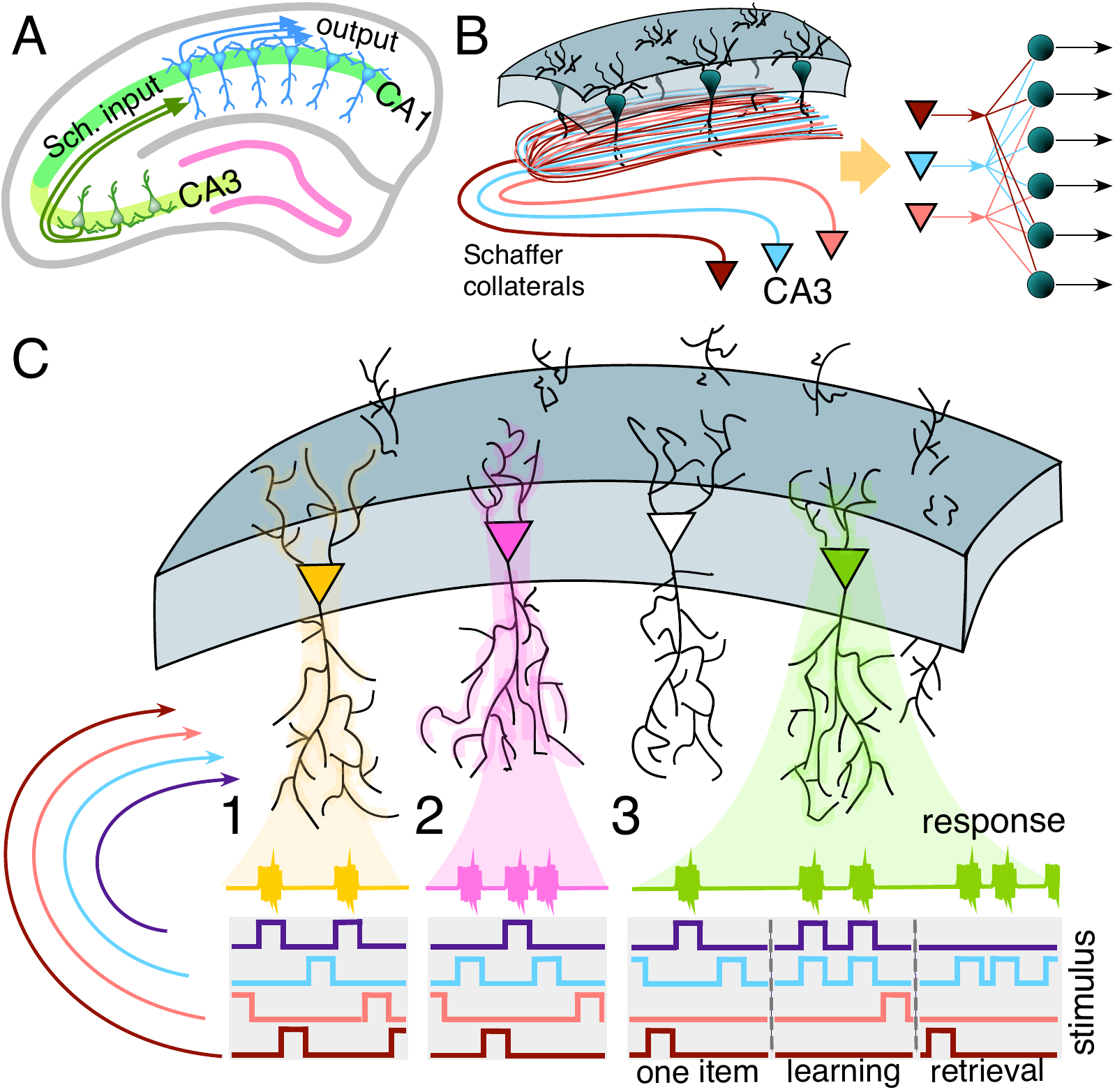, width = 0.85\textwidth}}
\caption{General principles of encoding memories by single neurons in laminar structures. A) Laminar organization of the CA3 and CA1 areas in the hippocampus facilitates multiple parallel synaptic contacts between neurons in these areas by means of Schaffer collaterals. B) Axons from CA3 pyramidal neurons bifurcate and pass through the CA1 area in parallel (left panel) giving rise to the convergence-divergence of the information content (right panel). Multiple CA1 neurons receive multiple synaptic contacts from CA3 neurons. C) Schematic representation of three memory encoding schemes. 1) Selectivity. A neuron ({shown} in yellow) receives inputs from multiple presynaptic cells that code different information items. It detects (responds to) only one stimulus (purple trace), whereas rejecting the others. 2) Clustering. Similar to 1, but now a neuron ({shown} in pink) detects a group of stimuli (purple and blue traces) and ignores the others. 3) Acquiring memories. A neuron ({shown} in green) learns dynamically a new memory item (blue trace) by associating it with a know one (purple trace). }
\label{Fig1}
\end{figure}

Different brain structures, such as, e.g., the hippocampus, have a pronounced laminar organization. For example the CA1 region of the hippocampus is constituted by a palisade of morphologically similar pyramidal cells oriented with their main axis in parallel and forming a monolayer (Fig. \ref{Fig1}A). The major excitatory input to these neurons comes through Schaffer collaterals from the CA3 region \cite{Amaral1989,Ishizuka1990,Wittner2007}, which is a hub routing information among many brain structures. Each CA3 pyramidal neuron sends an axon that bifurcates and leaves multiple collaterals in the CA1 with dominant parallel orientation (Fig. \ref{Fig1}B). This topology allows multiple parallel axons conveying multidimensional ``spatial'' information from one area (CA3) simultaneously leave synaptic contacts on multiple neurons in another area (CA1). Thus, we have simultaneous convergence and divergence of the information content (Fig. \ref{Fig1}B, right).

Experimental findings  show that multiple CA1 pyramidal cells distributed in the rostro-caudal direction are activated near-synchronously by assemblies of simultaneously firing CA3 pyramidal cells  \cite{Ishizuka1990,Li1994,Benito2014}. Thus, an ensemble of single neurons in the CA1 can receive simultaneously the same synaptic input (Fig. \ref{Fig1}B, left). Since these neurons have different topology and functional connectivity \cite{Rnnerty1983}, their response to the same input can be different. Moreover, experimental \textit{in-vivo} results show that long term potentiation can significantly increase the spike transfer rate in the CA3-CA1 pathway \cite{Fernandez2012}. This suggests that the efficiency of individual  synaptic contacts can be increased selectively.

In this work we will follow conventional and rather general functional representation of signalling in the neuronal pathways. We assume that upon  receiving an input, a neuron can either generate a response or remain silent. Forms of the neuronal responses as well as the definitions of synaptic inputs vary from one model to another. Therefore, here we adopt a rather general functional approach. Under a stimulus we understand a number of excitations simultaneously (or within a short time window) arriving to a neuron through several axones and thus transmitting some ``spatially coded'' information items \cite{Benito2016}. If a neuron responds to a stimulus (e.g., generates output spikes or increases its firing rate), we then say that the neuron \textit{detects} the informational content of the given  stimulus.

{We follow the standard machine learning assumptions \cite{Vapnik:2000}, \cite{cucker2002mathematical}. The stimuli are generated in accordance with some distribution or a set of distributions (``Outer World Models''). All stimuli that a neuron may receive are samples from this distribution. The sampling itself may be a complicated process, and for simplicity we assume that all samples are identically and independently distributed (i.i.d.). Once a sample is generated, a stimuli sub-sample is indepently selected for testing purposes. If more than one neuron is considered, we will assume that  a rule (or a set of rules) is in place that determines how a neuron is selected from the set. The rules can be both deterministic and randomized. In the latter case we will specify this process.}

Let us now pose the following fundamental questions related to the information encoding and formation of memories by  single neurons and their ensembles in laminated brain structures:

\begin{enumerate}
\item {\it Selectivity: Detection of one stimulus from a set} (Fig. \ref{Fig1}C.1). Pick an arbitrary stimulus from a reasonably large set such that a single neuron from a neuronal ensemble detects this stimulus. Then what is the probability that this neuron is stimulus-specific, i.e., it  rejects all the other stimuli from the set?
\vspace{0.2cm}

\item {\it Clustering: Detection of a group of stimuli from a set} (Fig. \ref{Fig1}C.2). Within a set of stimuli we select a smaller subset, i.e., a group of stimuli. Then what is the probability that a neuron detecting all stimuli from this subset stays silent for all remaining stimuli in the set?
\vspace{0.2cm}

\item {\it Acquiring memories: Learning new stimulus by associating it with one already known} (Fig. \ref{Fig1}C.3). Let us consider two different stimuli $\bfs_1$ and $\bfs_2$ such that {for $t\leq t_0$} they do not overlap in time and a neuron detects  $\bfs_1$, but not $\bfs_2$. In the next interval $(t_0,t_1]$, {$t_1>t_0$} the stimuli start to overlap in time (i.e., they stimulate the neuron together). {For} $t> t_1$ the neuron receives only  stimulus $\bfs_2$. Then what is the probability that for some $t_2\ge t_1$ the neuron detects $\bfs_2$?
\end{enumerate}

These questions are in the core of a broad range of puzzling phenomena reported in \cite{Quiroga:2015,Quiroga:2005,Quiroga:2009}. In what follows we will show that, remarkably, these {three non-trivial fundamental} questions can be answered within a simple  classical modeling framework, whereby a neuron is represented by a mere perceptron equipped with a Hebbian-type of learning.

\section{Formal statement of the problem}\label{sec:model_neuron}

In this section we  specify the information content to be processed by neurons and {define} a mathematical model of {a generic} neuron equipped with synaptic plasticity. Before going {any} further' let us first introduce {notational agreements used throughout the text}. Given two vectors  $\bfx,\boldsymbol{y}\in \Real^n$, their inner product {$\langle \bfx, \boldsymbol{y} \rangle$} is:
$\langle \bfx, \boldsymbol{y} \rangle=\sum_{i=1}^n x_i y_i$. {If $\bfx\in\Real^n$ then $\|\bfx\|$ stands for the usual Euclidean norm of $\bfx$: $\|\bfx\|=\langle\bfx,\bfx\rangle^{1/2}$. By  $B_n(1)=\{\boldsymbol{x}\in\Real^n | \ \ \|\bfx \|\leq 1\}$ we denote} a unit $n$-ball centered at the origin; $\Volume(\Xi)$ is the Lebesgue volume of  $\Xi \subset \Real^n$, and $|\mathcal{M}|$ is the cardinality of a finite set $\mathcal{M}$. {Symbol $\mathcal{C}(\mathcal{D})$, $\mathcal{D}\subseteq\Real^m$ stands for the space of continuous real-valued functions on $\mathcal{D}$}.

\subsection{Information content and classes of stimuli}

We assume that a neuron receives and processes a large but finite set of different stimuli codifying different information items:
\begin{equation}
\label{StimSet}
\mathcal{S}=\{ \bfs_i \}.
\end{equation}

\begin{figure}[!h]
\centering{\epsfig{file = 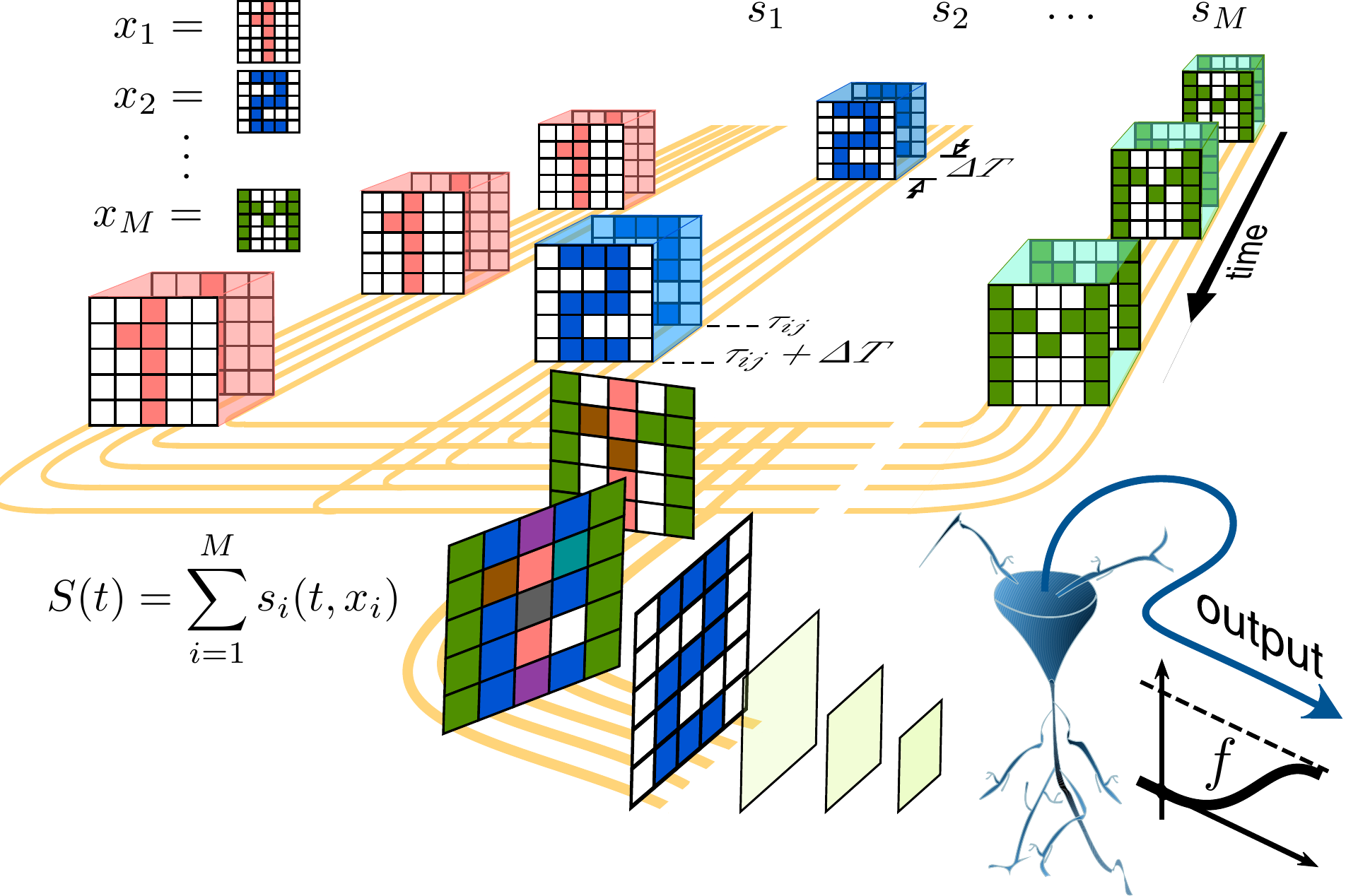, width = 0.85\textwidth}}
\caption{Codification of high-dimensional information by a neuron. Each of $M$ stimuli {comprises of} the ``spatial'' information, $\bfx_i\in \Real^n$, {(e.g., $M$ images)} conducted through $n$ axons (in yellow) and the temporal part, $c(t-{\tau_{i,j}})$, {reflecting} the times of {stimuli presentation}. A neuron (in blue) receives the stimuli and generates responses {determined} by some transfer function $f$.}
\label{Fig2}
\end{figure}

Figure \ref{Fig2} illustrates schematically the information flow. Each individual stimulus $i$ is modeled by a function $\boldsymbol{s}:\Real\times\Real^n\rightarrow \Real^n$:
\begin{equation}
\label{eq:stimulus_definition}
\bfs(t,\bfx_i)=\bfx_i \sum_{j}c(t-{\tau_{i,j}}),
\end{equation}
where $\bfx_i \in \Real^n\setminus \{0\}$ is the stimulus content codifying the information to be transmitted over $n$ individual ``axons''. {An example of an information item could be an $l\times k$  image (see Fig. \ref{Fig2}). In this case the dimension of each information item is $n = l\times k$.}

In Eq. (\ref{eq:stimulus_definition}) the function $c(\cdot)$ defines the stimulus \textit{context}, i.e., the time window when the stimulus arrives to the neuron. For the sake of simplicity we use a rectangular window:
\begin{equation}
\label{eq:spike_shape_definition}
c(t)=\left\{
\begin{array}{ll}
1, & \mbox{ if } t\in[0,\Delta T]  \\
0, & \mbox{ otherwise},
\end{array} \right .
\end{equation}
where $\Delta T > 0$ is the window length.  The time instants of the stimulus presentations, $\tau_{i,j}$, are ordered and satisfy:
\begin{equation}
\label{eq:spike_timings}
\tau_{i,j+1} > \tau_{i,j} + \Delta T,  \ \ \forall j.
\end{equation}

Different stimuli arriving to the neuron are added linearly on the neuronal membrane. Thus, the {overall} neuronal input {$\boldsymbol{S}$} can be written as:
\begin{equation}
\label{eq:NeurInput}
\boldsymbol{S}(t)  = \sum_{i,j} \boldsymbol{x}_i c(t-\tau_{i,j}).
\end{equation}

We assume that the  information content of stimuli (\ref{eq:NeurInput}) and (\ref{eq:stimulus_definition}), i.e., vectors $\bfx_i$ are drawn i.i.d. from some distribution. For convenience, we partition all information items into two sets:
\begin{equation}
\label{eq:M_set}
\mathcal{M}=\{\bfx_{1},\dots, \bfx_{M}\}, \ \ \mathcal{Y}=\{\bfx_{M+1},\dots, \bfx_{M+m}\},
\end{equation}
where $M$ is large but finite and $m\geq 1$  is in general smaller than $M$.  The set $\mathcal{M}$ contains a {\it background} content for a given neuron, whereas the set $\mathcal{Y}$ models the informational content {\it relevant} to the task at hand. In other words, to accomplish a static memory task the neuron should be able to detect all elements from $\mathcal{Y}$ and to reject all elements from $\mathcal{M}$.

The sets $\mathcal{M}$ and $\mathcal{Y}$ give rise to the corresponding subsets of stimuli:
\begin{equation}\label{eq:stimuli_M}
\begin{split}
\mathcal{S}(\mathcal{M})=\{\bfs_i\in\mathcal{S} \ | \  \bfs_i(\cdot)=\bfs(\cdot,\bfx_i), \ \bfx_i\in\mathcal{M} \}, \\
\mathcal{S}(\mathcal{Y})=\{\bfs_i\in\mathcal{S} \ | \  \bfs_i(\cdot)=\bfs(\cdot,\bfx_i), \ \bfx_i\in\mathcal{Y} \}.
\end{split}
\end{equation}

\subsection{Neuronal model}
\label{sect:model_neuron}

To stay within functional description of the information processing let us consider the most basic class of model neurons, a perceptron \cite{Rosenblatt1962}. A single neuron receives a stimulus $\boldsymbol{s}(t,\boldsymbol{x})$ through $n$ synaptic inputs (Fig. \ref{Fig2}) and its membrane potential, $y \in \Real$, is given by
\begin{equation}
\label{MembranePotential}
y(\bfs,\bfw)=\langle \bfw,\bfs \rangle,
\end{equation}
where $\bfw\in \Real^n$ is a vector of the synaptic weights. The
neuron generates a response, $v\in \Real$, according to:
\begin{equation}
\label{eq:neuron_model}
v(\bfs,\bfw,\theta)=  f(y(\bfs,\bfw) - \theta),
\end{equation}
where $\theta\in\Real$ is the ``firing'' threshold and $f:\Real\rightarrow\Real$ is the transfer function (Fig. \ref{Fig2}): $f \in \mathcal{C}(\Real)$, $f$ is locally Lipschitz, $f(u)=0$ for $u\in (-\infty,0]$, and $f(u)>0$ for $u\in(0,\infty)$.

Model (\ref{MembranePotential}), (\ref{eq:neuron_model}) captures the summation of postsynaptic potentials and the threshold nature of the neuronal activation but disregards the specific dynamics accounted for in other more advanced models. Nevertheless, as we will show in Sect. \ref{sec:results}, this phenomenological model is already sufficient to explain the fundamental properties of information processing discussed in Sect. \ref{sec:problem}.

\subsection{Synaptic plasticity}\label{sec:learning}

In addition to the basic neuronal response mechanism (Sect. \ref{sect:model_neuron}), we also model the synaptic plasticity.
The description adopted here relies on the neuronal firing rate {and Hebbian learning. Such a learning} rule implies that the dynamics of $\bfw$ should depend on the product of the input signal, $\bfs$, and the neuronal output, $v$. We thus arrive to a modified classical Oja rule \cite{Oja1982}:
\begin{equation}\label{eq:hebbian_oja}
\begin{split}
&\dot{\bfw}=\alpha  v(\bfs,\bfw,\theta) y(\bfs,\bfw) \left(\bfs  -  \bfw y(\bfs,\bfw) \right),\\
&\bfw(t_0)=\bfw_0\in\Real^n, \ \bfw_0\neq 0,
\end{split}
\end{equation}
where $\alpha>0$ defines the relaxation time. {The  multiplicative term $v$ in (\ref{eq:hebbian_oja}) ensures that plastic changes of $\bfw$ occur only when {an input stimulus} evokes a non-zero neuronal response. The fact that $\bfw_0\neq 0$ reflects the assumption that synaptic connections have already been established, albeit their efficacy could be subjected to plastic changes.} {In addition to capturing general principle of the classical Hebbian rule, model (\ref{eq:hebbian_oja}) guarantees that synaptic weights $\bfw$}  are bounded in forward time (see Appendix A) and hence conforms with physiological plausibility.

\section{Formation of memories in high-dimensions}
\label{sec:results}
In Sect. \ref{sec:problem} we formulated three fundamental problems of organization of memories in laminar brain structures. Let us now show how they can be treated given that pyramidal neurons operate in high dimensions.

To formalize the analysis let $\mathcal{U}$ be a subset of the stimulus set $\mathcal{S}$. A neuron (\ref{MembranePotential}), (\ref{eq:neuron_model}) parameterized by $(\bfw, \theta)$ partitions the set  $\mathcal{U}$ into the following subsets:
\begin{equation}
\begin{split}
\mathrm{Activated}(\mathcal{U},(\bfw,\theta))=&\{\bfs_i\in\mathcal{U} \ | \ \ \exists \, t{\geq} t_0: \ v(\bfs_i(t),\bfw,\theta) > 0\},\\
\mathrm{Silent}(\mathcal{U},(\bfw,\theta))=&\{\bfs_i\in\mathcal{U} \ | \ \ v(\bfs_i(t),\bfw,\theta)= 0 \ \ \forall \ t\geq t_0 \}.
\end{split}
\end{equation}
The first set corresponds to the stimuli detected by the neuron, while the second one collects background stimuli.

\subsection{Extreme selectivity of a single neuron to single stimuli}
\label{Section4.1}

{Consider the case when the set $\mathcal{Y}$ in (\ref{eq:M_set})} contains  only one element, i.e. $|\mathcal{Y}|=1$, {$\mathcal{Y}=\{\bfx_{M+1}\}$}, whereas  the set $\mathcal{M}$ is allowed to be sufficiently large ($|\mathcal{M}|=M\gg 1$). Let us also assume that the stimuli with different information content, $\bfs(\cdot,\bfx_i)$, do not overlap in time, i.e., we present them to a neuron one by one.

\begin{figure}[!h]
\centering{\epsfig{file = 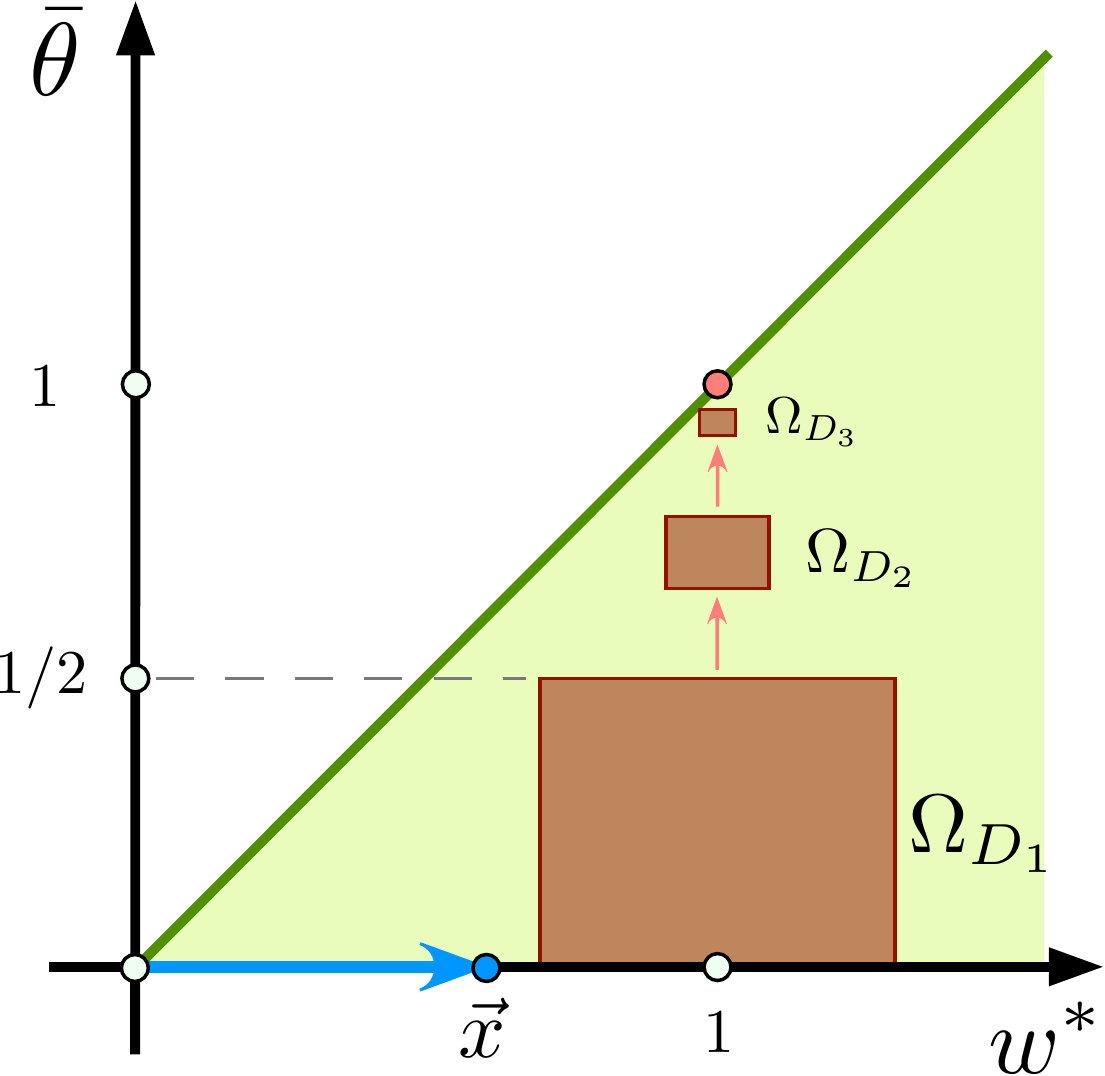, width = 0.4\textwidth}}
\caption{Selection of neuronal parameters $\bar{\theta}=\theta/\|\bfx_{M+1}\|$ and ${w^{\ast}}$, such that the neuron responds to the relevant information $\bfx_{M+1}$. {Neurons corresponding} to points within the green area detect the stimulus $\bfx_{M+1}$. Brown areas {show} projections of hypercylinders defined in Theorem \ref{theorem:selectivity} for $D_1 = 0.3$, $D_2$ = 0.1, $D_3 = 0.03$ and $\|\bfx_{M+1}\| = 0.6$.}
\label{Fig3}
\end{figure}

{For a given non-zero $\bfx_{M+1}\in \mathcal{Y}$ and stimulus $\bfs(\cdot,\bfx_{M+1})$ such that it is not identically zero for $t\geq t_0$ we can always construct a neuron  which would generate a non-zero response to the stimulus $\bfs(\cdot,\bfx_{M+1})$ at some $t\geq t_0$.} In other words, $\bfs(\cdot,\bfx_{M+1})\in\mathrm{Activated}(\mathcal{S}(\mathcal{Y}),(\bfw,\theta))$. Mathematically such a neuron can be defined as follows. {Let}
\begin{equation}
\label{eq:OrthonormalBasis}
\bfw^{\ast} = \frac{\bfx_{M+1}}{\|\bfx_{M+1}\|}.
\end{equation}
{Then the space from which the synaptic weights are chosen can be represented as a direct sum of the one-dimensional linear subspace $L^{\|}(\bfw^\ast)$  spanned by $\bfw^{\ast}$ and an $(n-1)$-dimensional subspace $L^{\bot}(\bfw^{\ast})$ of $\Real^n$ that is orthogonal to $\bfw^{\ast}$. In this representation, if a neuron with the synaptic  weight $\bfw$ generates a non-zero response to $\bfs(\cdot,\bfx_{M+1})$, then the coupling weight  $w^{\ast}=\langle\bfw,\bfw^{\ast}\rangle$ } should satisfy the following condition (Fig. \ref{Fig3}, green area):
\[
{w^{\ast}} > \frac{\theta}{\|\bfx_{M+1}\|}.
\]
{Indeed,} such a choice is {equivalent} to
\[
v(\bfx_{M+1},\bfw,\theta) = f({w^{\ast}}\|\bfx_{M+1}\| - \theta) > 0,
\]
{which in turn implies that $v(\bfs(t,\bfx_{M+1}),\bfw^\ast,\theta)>0$ at some $t$ and vice-versa.}

{Once a neuron that detects relevant information item, i.e. $\bfx_{M+1}$, is specified we can proceed with assessing} its selectivity properties.

\begin{definition}[Neuronal Selectivity]
\label{Def_Selective}
We say that a \textit{neuron is selective to the information content {$\mathcal{Y}$}} iff it detects the relevant stimuli from the set $\mathcal{S}(\mathcal{Y})$ and ignores all the others from the set $\mathcal{S}(\mathcal{M})$.
\end{definition}

{The notion of selectivity, as stated in Definition \ref{Def_Selective}, could be relaxed to account for partial detection and rejection of information content from $\mathcal{Y}$ and $\mathcal{M}$, respectively. This naturally gives rise to various levels of neuronal selectivity determined, for instance, by the proportion of elements from $\mathcal{M}$ that correspond to stimuli that have been rejected. As we will see below, different admissible pairs $(\bfw,\theta)$ (Fig. \ref{Fig3}) produce different selectivity levels. The closer to the bisector, the higher the selectivity. One can pick an arbitrary firing threshold $\theta \ge 0$ and select the synaptic efficiency at $t=t_0$ as:{
\begin{equation}
\label{eq:CouplWeights0}
\bfw(t_0) = \frac{\theta + \epsilon}{\|\bfx_{M+1}\|}\bfw^{\ast} + \bfw^{\bot}, \ \ \ \epsilon >0, \ \bfw^{\bot}\in L^{\bot}.
\end{equation}
It can be shown (see Appendix \ref{AppendixDynamicsW}) that if the stimulus $\bfs(\cdot,\bfx_{M+1})$  is persistent over time and $\bfw(t_0)$ satisfies (\ref{eq:CouplWeights0}) then synaptic efficiency $\bfw(t,\bfw_0)$ converges asymptotically (as $t\rightarrow\infty$) to:
\begin{equation}
\bfw_{\infty} =  \left \{
\begin{array}{ll}
\bfw^{\ast}, \ & \ \mbox{ if } \theta < \|\bfx_{M+1}\|\\
\frac{\displaystyle \theta}{\displaystyle \|\bfx_{M+1}\|}\bfw^{\ast} + \bfw^{\bot}_{\infty}, \ & \ \mbox{ if } \theta \ge \|\bfx_{M+1}\|,
\end{array}
\right.
\end{equation}
where $\bfw^{\bot}_{\infty}$ is an element of $ L^{\bot}$.}}

\begin{figure}[!h]
\centering{\epsfig{file = 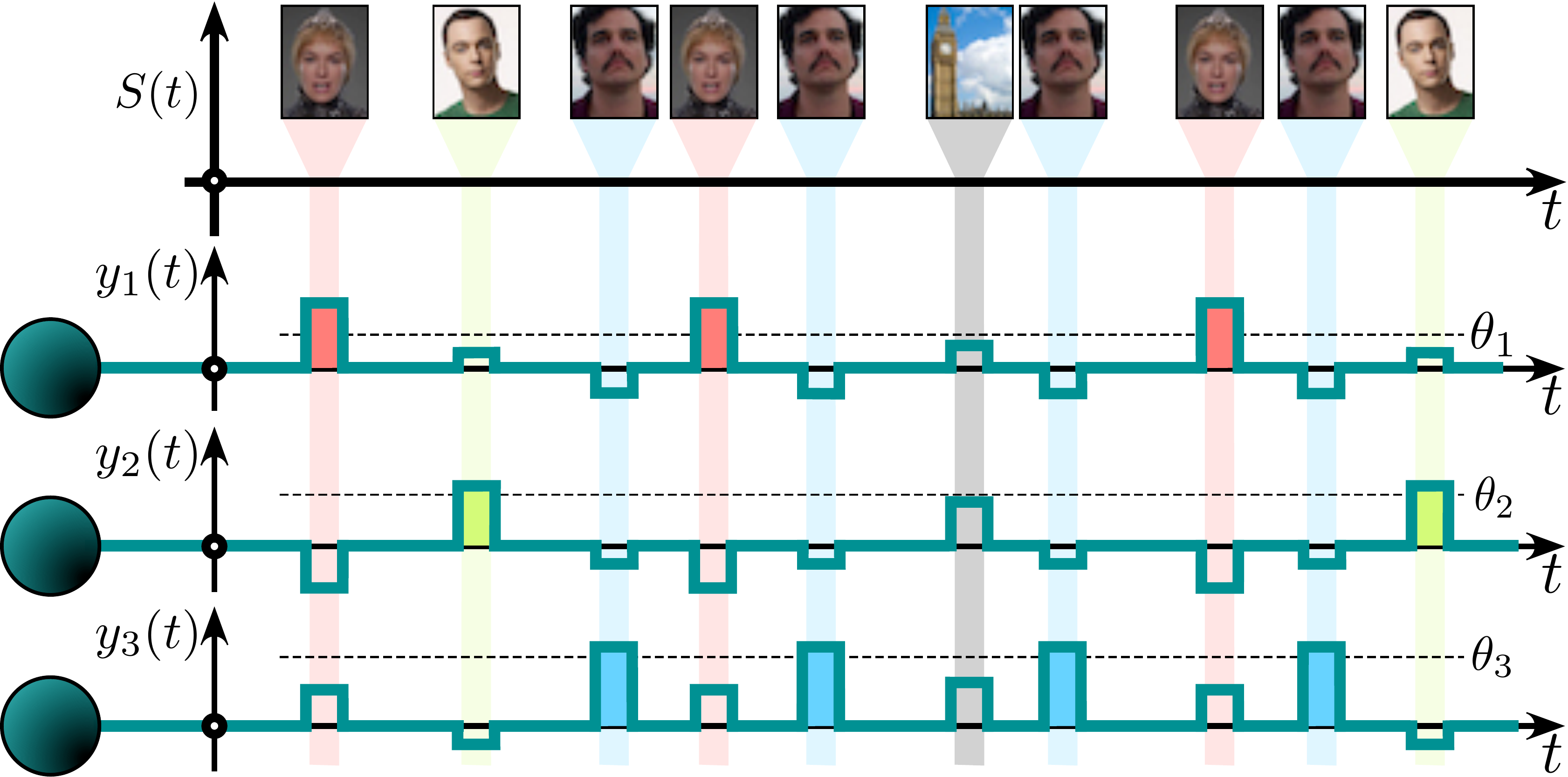, width = 0.95\textwidth}}
\caption{Example of selective neuronal responses to stimulation with different $(30\times 38)$-pixels images (only first few stimulus  are shown in the time line). Each neuron responds to its own (relevant) stimulus only and rejects the other (background) stimuli.}
\label{Fig4}
\end{figure}

{Figure \ref{Fig4} shows typical responses of neurons parameterized by} different pairs $(\bfw,\theta)$ {and subjected} to stimulation by different information items $\bfx_i$. {Here $\bfx_i$ correspond to} $(30 \times 38)$-pixels color images (i.e., $\bfx_i \in \Real^{3420}$). {Firing thresholds $\theta$ have been chosen at random, and weights $\bfw$ have been set in accordance with (\ref{eq:CouplWeights0}) with the first three images serving as the relevant information items for the three corresponding neurons. No plastic changes in $\bfw$ were allowed.} The neurons detect their own (relevant) stimuli, as expected. Moreover, they do not respond to the stimulation by other background information items (4 out of $10^3$ images are shown in Fig. \ref{Fig4}). Thus, the neurons indeed exhibit high stimulus selectivity.

The following theorem provides theoretical justification for these observations. 

\begin{theorem}
\label{theorem:selectivity} {Let elements of the sets $\mathcal{M}$ and $\mathcal{Y}$ be i.i.d. random vectors drawn from the equidistribution in $B_n(1)$.} Consider the sets of stimuli $\mathcal{S}(\mathcal{M})$ and $\mathcal{S}(\mathcal{Y})$ specified by (\ref{eq:stimuli_M}). Let $(\bfw,\theta)$ be the neuron parameters such that
\[
\bfs_{M+1}\in\mathrm{Activated}(\mathcal{S}({\mathcal{Y})},(\bfw,\theta)) \ \mbox{ and } \ 0<\theta<\|\bfw\|.
\]
Then:

\noindent
1. The probability that the neuron is silent for all background stimuli $\bfs_i\in\mathcal{S}(\mathcal{M})$ is bounded from below by:
\begin{equation}
\label{eq:selectivity_1}
\begin{split}
&P( \bfs_i \in \mathrm{Silent}(\mathcal{S}(\mathcal{M}),(\bfw,\theta)) \ \forall  \bfs_i\in\mathcal{S}(\mathcal{M}) \big| \ \bfw,\theta) \ge \\
& \ \ \ \ \ \ \ \ \geq
\left[
1-\frac{1}{2}
\left(1 - \frac{\theta^2}{\|\bfw\|^2}
\right)^\frac{n}{2}
\right]^M.
\end{split}
\end{equation}
2. There is a family of sets parametrized by $D$ ($0<D<\min\{\frac{1}{2}, \| \bfx_{M+1}\|\}$):
\begin{equation}
\label{OM_D}
\Omega_D=\Big \{ (\bfw,\theta)   \big| \ \ \|\bfw-\bfw^{\ast} \|<D, \  D \le \|\bfx_{M+1}\| - \theta \le 2D \Big \},
\end{equation}
where $\bfw^{\ast}=\bfx_{M+1}/\|\bfx_{M+1}\|$,  such that
$\bfs_{M+1}\in\mathrm{Activated}(\mathcal{S}(\mathcal{Y}),(\bfw,\theta))$, for $(\bfw,\theta)\in\Omega_D$ and
\begin{equation}
\label{eq:selectivity_2}
\begin{split}
&P\big( \bfs_i \in \mathrm{Silent}(\mathcal{S}(\mathcal{M}),(\bfw,\theta)) \ \forall \bfs_i\in\mathcal{S}(\mathcal{M})\big| \ \forall(\bfw,\theta)\in\Omega_D\big) \ge\\
& \ \ \ \ \ \ \ \ \ \ \geq \max_{\varepsilon\in(0,1-2D)} (1-(1-\varepsilon)^n) \left[1-\frac{1}{2} \rho(\varepsilon,D)^{\frac{n}{2}} \right]^M
\end{split}
\end{equation}
where
\[
\rho(\varepsilon,D)= 1 - \left(\frac{1-\varepsilon-2D}{1+D}\right)^2.
\]
\end{theorem}
The proof is provided in Appendix \ref{ProofTheorem1}.

\begin{remark} For {an admissible} fixed $D>0$,  the volume $\mathcal{V}(\Omega_D)>0$. Therefore, the estimate provided by Theorem \ref{theorem:selectivity} is robust to small perturbations of $(\bfw,\theta)$, and {slight fluctuations of neuronal characteristics are not expected to affect neuronal functionality.}
\end{remark}

\begin{remark}
Theorem \ref{theorem:selectivity} {(part 2) specifies a non-iterative procedure for constructing sets of selective neurons}. Such neurons detect given stimuli and reject the others, with high probability. Figure \ref{Fig3} (in brown) shows examples of three projections of the hypercylinders (\ref{OM_D}) ensuring robust selective stimulus detection. The smaller is the cylinder, the higher is the selectivity.
\end{remark}

To illustrate Theorem \ref{theorem:selectivity} {numerically} {we fixed the neuronal dimensionality parameter $n$ and generated two random sets of information items comprising of $10^3$ elements each, i.e. $\{\bfx_i\}_{i=1}^{10^3}$. One set was sampled from the equidistribution in a unit ball $B_n(1)$ centered at the origin (i.e. $\|\bfx_i\|_2 \le 1$), and the {other from the equidistribution in the hypercube $\|\bfx_i\|_{\infty} \le 1$ (a product distribution). For each set of informational items, a neuronal ensemble of $10^3$ single neurons {parameterized by $(\bfw_i,\theta_i)$ was created. Each neuron was assigned а fixed firing threshold $\theta_i = 0.5$, $i=1,\dots,10^3$, whereas the synaptic efficiencies were set as $\bfw_i=(\theta_i+\epsilon)\bfx_i/\|\bfx_i\|$, $\epsilon =0.05$}. For these neuronal ensembles and their corresponding stimuli sets  we evaluated output of each neuron and assessed the neuronal selectivity (see Def. \ref{Def_Selective})}. The procedure was repeated 10 times. This was followed by evaluation of the frequencies of selective neurons in the pool for each $n$.}

\begin{figure}[!h]
\centering{\epsfig{file = 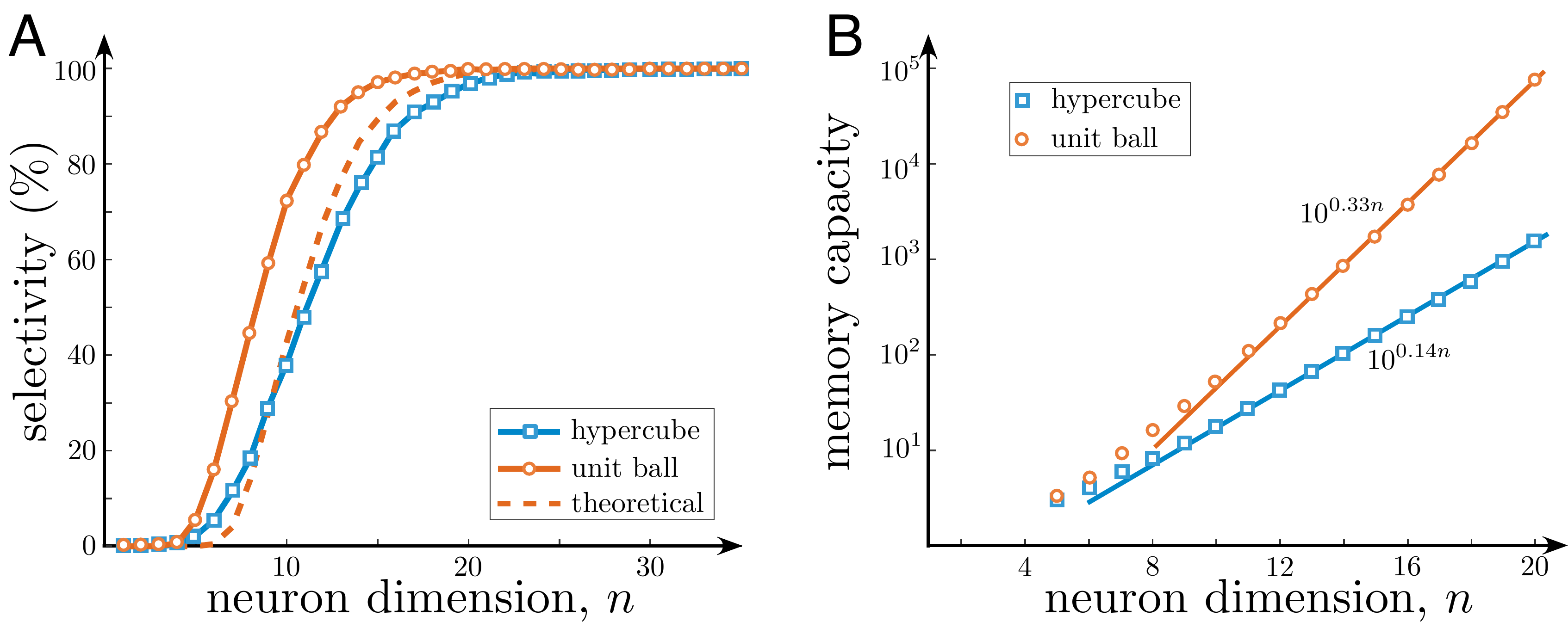, width = 0.95\textwidth}}
\caption{Extreme selectivity to stimuli and memory capacity of single neurons. A) Stimulus selectivity vs the neuron dimension. {The selectivity index steeply increases for $n\in[10,20]$. For $n>20$ practically all neurons become selective to a set of $10^3$ random stimuli.} B) Memory capacity {with reliability $0.95$ of a neuronal ensemble vs the neuron dimension. For both types of stimuli the memory capacity grows exponentially (straight lines show regressions).}}
\label{Fig5}
\end{figure}

Figure \ref{Fig5}A shows frequencies of selective neurons in an ensemble, for {$10^3$} stimuli taken from: i)  a unit ball (red), ii) a hypercube (blue), and iii) the estimate provided by Theorem \ref{theorem:selectivity} (dashed). For $n$ small ($n < 6$) neurons exhibit {no} selectivity, i.e., they confuse different  stimuli and generate nonspecific responses. As expected, when neuronal dimensionality, $n$, increases,  the neuronal selectivity  increases rapidly; and at around $n = 20$ it approaches $100\%$.

\subsection{Extreme selectivity of a single neuron and ensemble memory capacity}

{The property of a neuron to respond selectively to a single element from a large set of stimuli can be related to the notion of {\it memory capacity} of a neuronal ensemble  comprising of a set of selective neurons.

{Recall that in the framework of associative memory \cite{hopfield1987neural}, for each informational item (pattern) $\bfx_i$ from the set $\mathcal{M}$ there is a vicinity $\mathcal{V}_i$ associated with $\bfx_i$ and corresponding to all admissible perturbations of $\bfx_i$. Suppose that for each $\bfx_i$ there is a neuron in the ensemble that is activated for all stimuli with informational content $\bfx$ in $\mathcal{V}_i$  and is silent for all other stimuli, i.e. for stimuli with $\bfx$ in $\cup_{j\neq i}\mathcal{V}_j$. The maximal size of the set $\mathcal{M}$ for which this property holds will be referred to as  the {\it (absolute) memory capacity} of the ensemble (cf. \cite{hopfield1987neural}, \cite{barrett2004individual}, \cite{leung1995stability}).}

{This conventional mechanistic definition of memory capacity, however, is too restrictive to account for variability and uncertainty that biological neuronal ensembles and systems are to deal with. Indeed, informational items themselves may bear a degree of uncertainty resulting in that $\mathcal{V}_i\cap \mathcal{V}_j\neq \varnothing$ for some $j,i$, $i\neq j$. Furthermore, errors in memory retrievals are known to occur in classical artificial associative memory models too (see e.g., \cite{hopfield1987neural}, \cite{amit1987storing}, \cite{leung1995stability}). To be able to formally quantify such errors in relation to the number of informational items an ensemble is to store, we extend the classical notion as follows.}

{Suppose that for each $\bfx_i$ there is a neuron in the ensemble that is activated for all stimuli with informational content $\bfx\in\mathcal{V}_i$ and, with probability $\phi$, is silent for all stimuli with $\bfx\in \mathcal{V}_j$, $j\neq i$. The maximal size of the set $\mathcal{M}$ for which this property holds will be referred to as  the  {\it memory capacity with reliability $\phi$} of the ensemble.}

{Assuming that $\mathcal{V}_i$ are sufficiently small an estimate of the memory capacity with reliability $\phi$ of a neuronal ensemble follows from Theorem \ref{theorem:selectivity}}.

\begin{corollary}
\label{LEMMA_1} {Let elements of the sets $\mathcal{M}$ and $\mathcal{Y}$ be i.i.d. random vectors drawn from the equidistribution in $B_n(1)$.}
Consider the set of stimuli $\mathcal{S}(\mathcal{M})$ as defined in (\ref{eq:stimuli_M}). Then for a given fixed $\phi\in(0,1)$
the {maximal} size $\overline{M}$ of the stimuli set $\mathcal{S}(\mathcal{M})$ for which the following holds
\[
P( \bfs_i \in \mathrm{Silent}(\mathcal{S}(\mathcal{M}),(\bfw,\theta)) \ \forall  \bfs_i\in\mathcal{S}(\mathcal{M}) \big| \ \bfw,\theta) \ge \phi
\]
grows at least exponentially with the neuronal dimension $n$:
\begin{equation}
\label{LEM:1}
\overline{M} >  -\ln\left( \phi \right)\left(2e^{\alpha n} - 1\right), \  \mbox{ where } \
\alpha = \ln \left[\frac{\|\bfw\|}{\sqrt{\|\bfw\|^2-\theta^2}}\right] >0.
\end{equation}

\end{corollary}
The proof is given in Appendix \ref{ProofLEMMA_1}.

{Figure \ref{Fig5}B illustrates how the memory capacity with {reliability $\phi$ grows with neuronal dimension $n$}. For each neuronal dimension $n$ we generated i.i.d. samples $\mathcal{M}$ with $|\mathcal{M}|=M$ from the equidistribution in $B_n(1)$ and the $n$-cube $[-1,1]^n$. For each sample, we defined neuronal ensembles comprising of $M$ neurons with synaptic weights $\bfw_i=\bfx_i/\|\bfx_i\|$ and thresholds $\theta_i=0.5$, and calculated the proportion of neurons in the ensemble that are activated by each stimulus. If the proportion was smaller than $0.05$ of the total number of neurons, we incremented the value of $M$,  generated a new sample $\mathcal{M}$ with increased cardinality $M$, and repeated the experiment. The values of $M$ corresponding to samples at which the process stopped have been recorded and retained. These constituted empirical estimates of the maximal number of stimuli for which the proportion of neurons responding to a single stimulus} is at most $0.05=1-\phi$. Figure \ref{Fig5}B shows {empirical means} of such numbers for the unit ball and in the hypercube. As follows from these observations, memory capacity grows exponentially with the neuron dimension in both cases. Such a fast growth can easily cover quite exigent memory necessities.


\subsection{Selectivity of a single neuron to multiple stimuli}
\label{Section4.2}

{To organize memories, the ability to associate different information items is essential (Fig. \ref{Fig1}C2). To determine if such associations are feasible at the level of single neurons we assess neuronal selectivity to multiple stimuli. In particular, we  consider the set $\mathcal{Y}$ [Eq. (\ref{eq:M_set})] containing $m>1$ random vectors: $\mathcal{Y}=\{\bfx_{M+1},\dots, \bfx_{M+m}\}$. As in Sect. \ref{Section4.1}, here we assume that all stimuli do not overlap in time and arrive to the neuron separately. The question of interest is: Can we find a neuron [i.e., parameters $(\bfw,\theta)$], such that it would generate a non-zero response to all $\bfs_i\in \mathcal{S}(\mathcal{Y})$ and, with high enough probability, would be silent to all $\bfs_i \in\mathcal{S}(\mathcal{M})$?}

Below we will show that this is indeed possible, provided that the neuronal dimensionality, $n$, is large enough. Moreover, the separation can be achieved by a neuron with the vector of synaptic weights, {$\bfw=\bfw^\ast$},  closely aligned with the mean vector of the stimulus set $\mathcal{Y}$:
\begin{equation}
\label{eq:meanvect}
\bar{\bfx}=\frac{1}{m}\sum_{i=1}^{m} \bfx_{M+i}, \ {\bfw^\ast=\frac{\bar{\bfx}}{\|\bar{\bfx}\|}}.
\end{equation}
This vector points to the center of the group to be separated from the set $\mathcal{M}$. In low dimensions, e.g. when  $n = 2$, such  functionality appears to be extremely unlikely. However, high dimensional neurons can accomplish this task with probability close to one. {Formal statement of this property is provided in Theorem \ref{theorem:selectivity_multiple}.}
\begin{theorem}
\label{theorem:selectivity_multiple} {Let elements of the sets $\mathcal{M}$ and $\mathcal{Y}$ be i.i.d. random vectors drawn from the equidistribution in $B_n(1)$.} Consider the sets of stimuli $\mathcal{S}(\mathcal{M})$ and $\mathcal{S}(\mathcal{Y})$ specified by (\ref{eq:stimuli_M}) and let $D, \, \varepsilon, \, \delta\in(0,1)$ be chosen such that
\begin{equation}\label{eq:k-tuple:functional:neuron}
\theta^\ast=\frac{(1-\varepsilon)^3 - \delta(m-1)}{\sqrt{m(1-\varepsilon )[1-\varepsilon + \delta(m-1)]}} \in (D,1).
\end{equation}
{Let $\bfw^{\ast}  = \bar{\bfx}/\| \bar{\bfx} \|$} and consider the set:
\[
\Omega_D=\Big\{(\bfw,\theta) \big| \ \|\bfw-\bfw^{\ast}\|< D, \  \theta  \in (0, \theta^\ast-D] \Big\}.
\]
Then
\begin{equation}\label{eq:prob_selectivity2}
\begin{split}
&P\Big( [ \bfs_i\in \mathrm{Activated} (\mathcal{S}(\mathcal{Y}),\bfw,\theta) \ \forall  \ \bfs_i\in\mathcal{S}(\mathcal{Y})\mathcal ] \  \& \ \\
 & \ \  [ \bfs_i\in \mathrm{Silent} (\mathcal{S}(\mathcal{M}),\bfw,\theta) \ \forall \ \bfs_i\in\mathcal{S}(\mathcal{M})] \Big | \ (\bfw,\theta)\in\Omega_D \Big)\geq  p(\varepsilon,\delta,D,m),
\end{split}
\end{equation}
where
\[
\begin{split}
p(\varepsilon,\delta,D,m)=& (1-(1-\varepsilon)^n)^{m}\prod_{d=1}^{m-1} \left(1-d \left(1 -\delta^2\right)^{\frac{n}{2}} \right) \left[1 -
\frac{1}{2}\Delta^\frac{n}{2}\right]^{M}, \\
&\Delta=1- \frac{\theta^2}{(1+D)^2}.
\end{split}
\]
\end{theorem}
The proof is provided in Appendix \ref{ProofTheorem2}. The theorem admits {the following corollary.}

\begin{corollary} \label{cor:selectivity_multiple}  Suppose that the conditions of Theorem \ref{theorem:selectivity_multiple} hold. Let $\theta^\ast>2D$ and consider the set:
\[
\Omega_D^\ast=\Big\{(\bfw,\theta) \big| \ \|\bfw-\bfw^{\ast}\|< D, \  \theta  \in [\theta^\ast - 2D, \theta^\ast-D] \Big\}.
\]
Then
\begin{equation}\label{eq:prob_selectivity3}
\begin{split}
&P\Big( [ \bfs_i\in \mathrm{Activated} (\mathcal{S}(\mathcal{Y}),\bfw,\theta) \ \forall  \ \bfs_i\in\mathcal{S}(\mathcal{Y})\mathcal ] \  \& \ \\
 &  \  [ \bfs_i\in \mathrm{Silent} (\mathcal{S}(\mathcal{M}),\bfw,\theta) \ \forall \ \bfs_i\in\mathcal{S}(\mathcal{M})] \Big | (\bfw,\theta)\in\Omega_D^\ast \Big)\geq  \\
& (1-(1-\varepsilon)^n)^{m}\prod_{d=1}^{m-1} \left(1-d \left(1 -\delta^2\right)^{\frac{n}{2}} \right) \left[1 -
\frac{1}{2}\Delta^\frac{n}{2}\right]^{M}, \\
&\Delta=1- \left(\frac{\theta^\ast -2D}{1+D}\right)^2.
\end{split}
\end{equation}
\end{corollary}

\begin{remark} Estimates {(\ref{eq:prob_selectivity2}), (\ref{eq:prob_selectivity3}) hold for all feasible values of $\varepsilon$ and $\delta$. Maximizing the r.h.s of (\ref{eq:prob_selectivity2}), (\ref{eq:prob_selectivity3}) over feasible domain of $\varepsilon$, $\delta$ provides lower-bound ``optimistic'' estimates of the neuron performance.}
\end{remark}

\begin{remark}
\label{RemarkDecay}
The term $\theta^\ast$ in Theorem \ref{theorem:selectivity_multiple} and {Corollary \ref{cor:selectivity_multiple}} is an upper bound for the firing threshold $\theta$. The larger is the value of $\theta$, the higher is the neuronal selectivity to multiple stimuli. The value of $\theta^\ast$, however, decays with the number of stimuli  $m$.
\end{remark}
The extent to which the decay mentioned in Remark \ref{RemarkDecay} affects neuronal selectivity to a group of stimuli depends largely on the neuronal dimension, $n$. Note also that the probability of neuronal selective response to multiple stimuli, as provided by Theorem \ref{theorem:selectivity_multiple}, can be much larger if elements of the set $\mathcal{Y}$ are spatially close to each other or positively correlated \cite{TyuGor2017knowlege} (see also Lemma \ref{lem:k-tuples:ball:correlated} in Appendix \ref{AppendixAux}).

\begin{remark}\label{rem:exponential:2} Similarly to the case considered in Corollary \ref{LEMMA_1}, the {maximal} size of the stimuli set $\mathcal{S}(\mathcal{M})$ for which selective response is ensured, with some fixed probability, grows exponentially with dimension $n$. {Indeed, denoting $\phi=(1-z)^{\overline{M}}$, letting $z=1/2 \Delta^{n/2}$ (with $\Delta$ defined in Theorem \ref{theorem:selectivity_multiple}) and invoking (\ref{po}), (\ref{APPC_1}) from the proof of Corollary \ref{LEMMA_1},  we observe}  that
\[
\overline{M}>-\ln (\phi)(z^{-1}-1)=-\ln (\phi) (2 e^{\beta n}-1), \ \beta=\ln \frac{1+D}{\sqrt{(1+D)^2-\theta^2}}.
\]
{Thus, for  $M=|\mathcal{S}(\mathcal{M})|\leq \overline{M}$, the r.h.s. of (\ref{eq:prob_selectivity2}) is bounded from below by
\[
(1-(1-\varepsilon)^n)^{m}\prod_{d=1}^{m-1} \left(1-d \left(1 -\delta^2\right)^{\frac{n}{2}} \right) \phi.
\]
Similar estimate can be provided for the case considered in Corollary \ref{cor:selectivity_multiple}.}
\end{remark}

\begin{figure}[!h]
\centering{\epsfig{file = 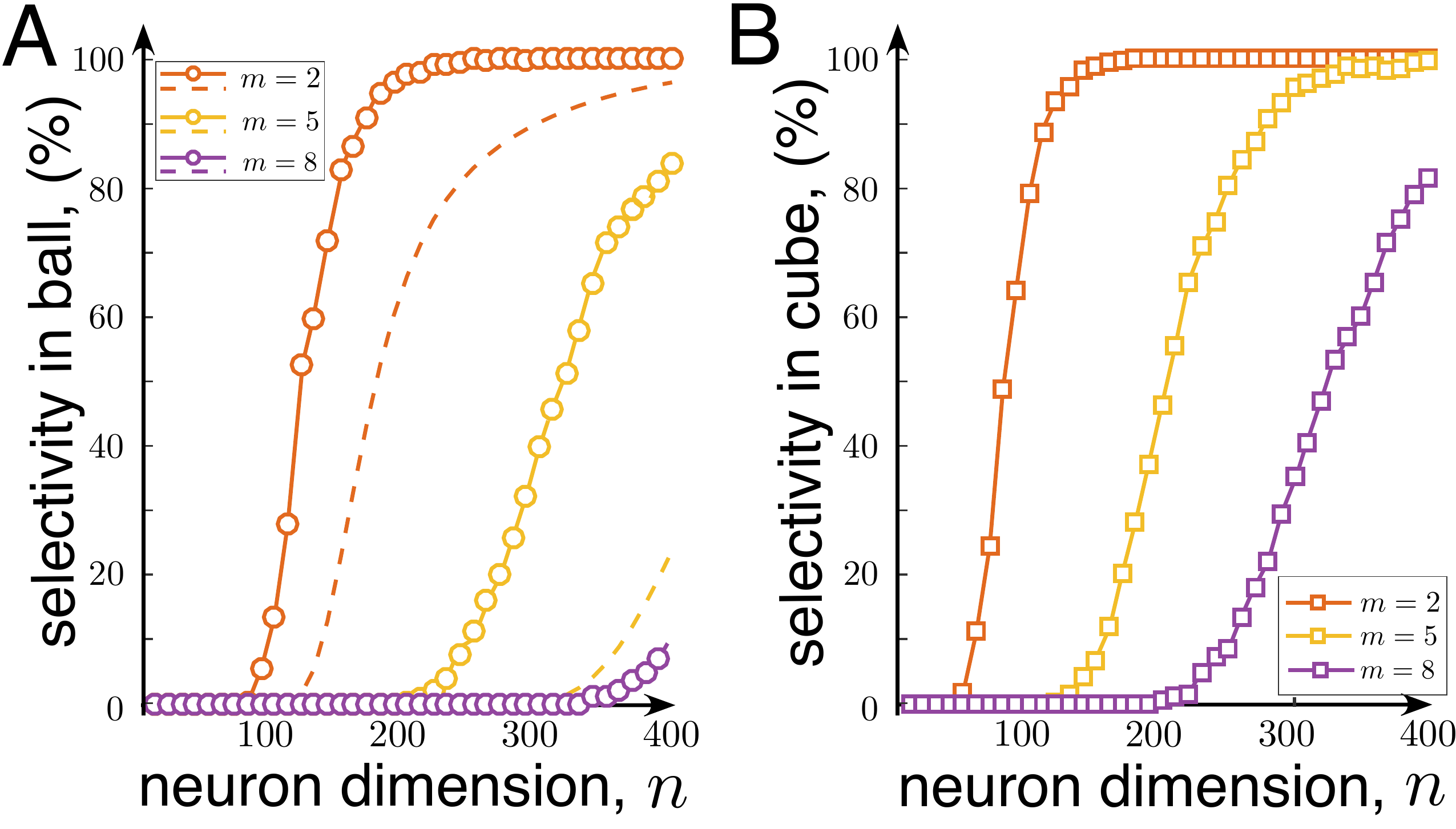, width = 0.85\textwidth}}
\caption{Selectivity of a single neuron to multiple stimuli. {Panel (A) corresponds to the case when the informational content vectors, $\bfx_i$,  are sampled from the equidistribution in the unit ball $B_n(1)$, and panel (B) corresponds to the equidistribution in the $n$-cube centered in the origin. In both cases the neuronal selectivity approaches $100\%$ when the dimension $n$ grows.} In (A) dashed curves show the estimates provided by Theorem \ref{theorem:selectivity_multiple}. Parameter values: $\varepsilon = 0.01$, $D = 0.001$, $\delta = (1 - \varepsilon)/2(m-1), \theta = \theta^* - D$.}
\label{Fig6}
\end{figure}

To illustrate Theorem \ref{theorem:selectivity_multiple} we conducted several numerical experiments. {For each $n$ we generated $M=10^3$ of background information items $\bfx_i$ (the set $\mathcal{M}$) and $m=2, 5, 8$ relevant vectors (the sets $\mathcal{Y}$)}. In the first group of experiments all $M+m$ i.i.d. random vectors were chosen from the equidistribution in $B_n(1)$. Neuronal parameters were set in accordance with Theorem \ref{theorem:selectivity_multiple} {(i.e., Eqs. (\ref{eq:meanvect}) -- (\ref{eq:prob_selectivity2}))}. Figure \ref{Fig6}A illustrates the results.

Similarly to the case of neuronal selectivity to a single item (Fig. \ref{Fig5}A), we observe a steep growth of the selectivity index with the neuronal dimension. The sharp increase occurs, however, at significantly higher dimensions. The number of random and uncorrelated stimuli, $m$, to which a neuron should be able to respond selectively is fundamentally linked to the neuron dimensionality. For example, the probability that a neuron is selective to $m=5$ random stimuli becomes sufficiently high only at $n > 400$. This contrasts sharply with $n=120$ for $m=2$.

Our numerical experiments also show that the firing threshold specified in Theorem \ref{theorem:selectivity_multiple} for arbitrarily chosen fixed values of $\delta$ and $\varepsilon$ is not optimal in the sense of providing the best possible probability estimates. Playing with $\theta$ one can observe that the values of $n$ at which neuronal selectivity to multiple stimuli starts to emerge are in fact significantly lower than those predicted by Eq. (\ref{eq:prob_selectivity3}). This is not surprising.  First, since estimate (\ref{eq:prob_selectivity3}) holds for all admissible values of $\delta$ and $\varepsilon$, it should also hold for the maximizer of $p(\varepsilon,\delta,D,m)$. Second, the estimate is conservative in the sense that it is based on conservative estimates of the volume of spherical cups $\mathcal{C}_n$ (see, e.g., proof of Theorem \ref{theorem:selectivity}). Deriving more accurate numerical expressions for the latter is possible, although at the expense of simplicity.

To demonstrate that dependence of the selectivity index on the firing threshold is likely to hold qualitatively for broader classes of distributions from which the sets $\mathcal{M}$ and $\mathcal{Y}$ are drawn, we repeated the simulation for the equidistribution in an $n$-cube centered at the origin. In this case, Theorem \ref{theorem:selectivity_multiple} does not formally apply. Yet, an equivalent statement can still be produced (cf. \cite{GorbanTyukin:NN:2017}). {In these experiments synaptic weights were set to $\bfw=\bar{\bfx}/\|\bar{\bfx}\|$} and  $\theta = 0.5\|\bar{\bfx}\|$. The results are shown in Fig. \ref{Fig6}B. The neuron's performance in the cube is markedly better than that of in $B_n(1)$. Interestingly, this is somewhat contrary to expectations that might have been induced by our earlier experiments (shown in Fig. \ref{Fig5}) in which neuronal selectivity  to a single stimulus was more pronounced for $B_n(1)$.

Overall, these results suggest that single neurons can indeed separate random uncorrelated information items from a large set of background items with probability close to one. This gives rise to a possibility for a neuron to respond selectively to various arbitrary uncorrelated information items simultaneously. The latter property provides a natural mechanism for accurate and precise grouping of stimuli in single neurons.

\subsection{Dynamic memory: Learning new information items by association}

In the previous sections we dealt with a static model of neuronal functions, i.e. when the synaptic efficiency $\bfw$ either did not change at all or the changes were negligibly small over {large intervals of stimuli presentation. In the presence of synaptic plasticity (\ref{eq:hebbian_oja}), the latter case corresponds to  $0\leq \alpha\ll 1$ in (\ref{eq:hebbian_oja})}. In this section we explicitly {account for} the time evolution of the synaptic efficiency, $\bfw(t,\bfw_0)$ [Eq. (\ref{eq:hebbian_oja})]. {As we will see below, this may give rise to dynamic memories in single neurons.}

As before, we will deal with two sets of stimuli, the relevant one, $\mathcal{S}(\mathcal{Y})$, and the background one, $\mathcal{S}(\mathcal{M})$. We will consider two time epochs: i) Learning phase and ii) Retrieval phase.  Within the learning phase {we assume that} all stimuli from the set $\mathcal{S}(\mathcal{Y})$ arrive to a neuron completely synchronized, i.e.:
\begin{equation}\label{eq:synchrony_y}
\tau_{M+1,j}=\tau_{M+2,j}=\cdots = \tau_{M+m,j}, \ \ \forall \ j.
\end{equation}
{Such a} synchronization {could be interpreted as}  a mechanism for associating or grouping different uncorrelated information items for the purposes of memorizing them at a later stage.

The dynamics of the synaptic weights for $t\geq t_0$ is given Eq. (\ref{eq:hebbian_oja}) with the input signal {$\bfs$ replaced} with:
\begin{equation}
\label{eq:hebbian_oja_sum}
\bar{\bfs}(t)= \sum_{i=1}^m \bfs_{M+i}(t).
\end{equation}
Let $\bfw_0={\bfw(t_0)}$ and $\theta$ satisfy the following condition:
\begin{equation}\label{eq:active_stimulus}
\begin{split}
& \exists \ \bfs_k\in\mathcal{S}(\mathcal{Y}) \ \mbox{ such that } \ \bfs_k\in\mathrm{Activated}(\mathcal{S}(\mathcal{Y}),\bfw_0,\theta) \\
&\bfs_i\in\mathrm{Silent}({\mathcal{S}},\bfw_0,\theta) \ \mbox{for all} \ \bfs_i\in {\mathcal{S}}\setminus \{\bfs_k\}.
\end{split}
\end{equation}
{Thus, at $t=t_0$ only one information item is} ``known'' to the neuron. All other relevant items {from the set $\mathcal{Y}$} are ``new'' {in the sense that} the neuron rejects them at $t=t_0$.  {Theorem \ref{theorem:selectivity} specifies the sets of neuronal parameters $\bfw_0,\theta$  for which condition (\ref{eq:active_stimulus}) holds with probability close to one if} $n$  is large enough.

The question is: What is the probability that, during the learning phase the synaptic weights $\bfw(t,\bfw_0)$ evolve in time {so} that the neuron becomes responsive to all $\bfs_i\in\mathcal{S}(\mathcal{Y})$ whilst remaining silent to all $\bfs_i\in\mathcal{S}(\mathcal{M})$ (Fig. \ref{Fig1}C.3)? In other words, the neuron learns new items and recognizes them in the retrieval phase. {The following theorem provides an answer to this question}.

\begin{theorem}
\label{theorem:new_memories} {Let elements of the sets $\mathcal{M}$ and $\mathcal{Y}$ be i.i.d. random vectors drawn from the equidistribution in $B_n(1)$.} Consider the sets of stimuli $\mathcal{S}(\mathcal{M})$ and $\mathcal{S}(\mathcal{Y})$ specified by (\ref{eq:stimuli_M}). Let (\ref{eq:synchrony_y}) hold, the dynamics of neuronal synaptic weights satisfy (\ref{eq:hebbian_oja}), (\ref{eq:hebbian_oja_sum}), and $(\bfw_0, \theta)$ be chosen such that condition (\ref{eq:active_stimulus}) is satisfied. Pick $\varepsilon,\delta\in(0,1)$  such that
\[
(1-\varepsilon)^3 > \delta(m-1).
\]
Moreover, suppose that
\begin{enumerate}
\item There exist $L,\kappa>0$ such that
\[
\int_{t}^{t+L} v(\bar{\bfs}(\tau),\bfw(\tau,{\bfw_0}),\theta) \langle \bar{\bfs}(\tau),\bfw(\tau,{\bfw_0}) \rangle^2 {d\tau} > \kappa, \ \ \forall \ t\geq t_0.
\]
\item  The firing threshold, $\theta$, satisfies
\[
0<\theta< \frac{(1-\varepsilon)^3 - \delta(m-1)}{\sqrt{m(1-\varepsilon)[(1-\varepsilon)+\delta(m-1)]}}={\theta^\ast}.
\]
\end{enumerate}
Then for, any {$0<D\leq\theta^\ast-\theta$}, there is  $t_1(D)> t_0$ such that
\[
\begin{split}
&P([\mathcal{S}(\mathcal{Y})\in\mathrm{Activated}(\mathcal{S},\bfw(t,{\bfw_0}),\theta)]\ \& \ [\mathcal{S}(\mathcal{M})\in\mathrm{Silent}(\mathcal{S},\bfw(t,{\bfw_0}),\theta)])\geq\\
& (1-(1-\varepsilon)^n)^{m}\prod_{d=1}^{m-1} \left(1-d \left(1 -\delta^2\right)^{\frac{n}{2}} \right) \left[1 - \frac{1}{2}\left(1-\frac{\theta^2}{(1+D)^2}\right)^\frac{n}{2}\right]^{M}
\end{split}
\]
for all $t\geq t_1(D)$.
\end{theorem}
The proof is provided in Appendix \ref{ProofTheorem3}.

\begin{figure}[!h]
\centering{\epsfig{file = 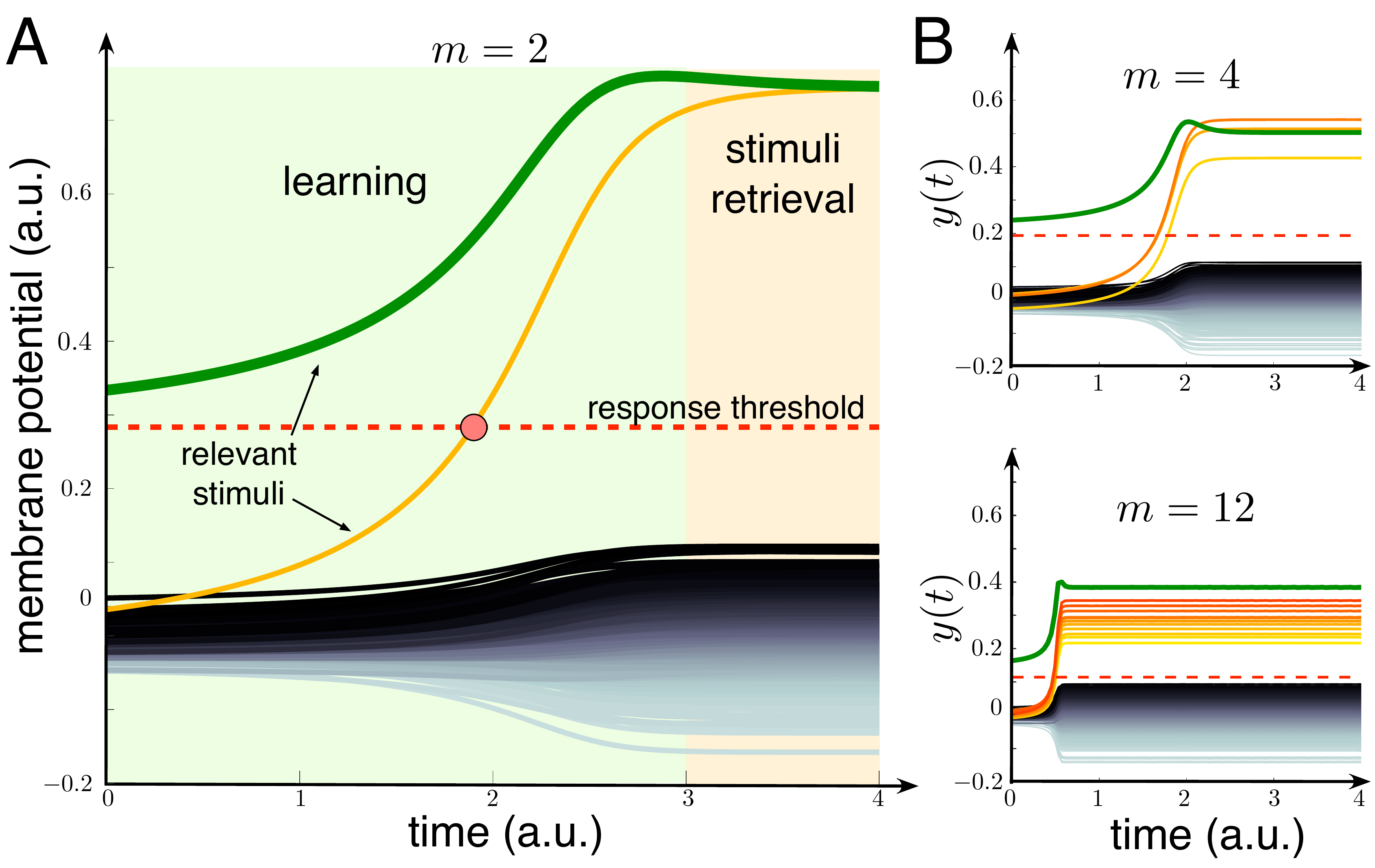, width = 1\textwidth}}
\caption{Dynamic memory: Learning new information items by {association}. A) Example of the dynamic association of a known stimulus ({neuron's response to the known stimulus is shown by} green curve) and a new one ({neuron's response shown by} orange curve). Two relevant stimuli out of 502 are learnt by the neuron. At $t\approx 2$ (red circle) the orange {curve} crosses the threshold (red dashed line) and {stays above it for $t>2$}. Thus the neuron detects {the corresponding stimulus for $t>2$}. B) Same as in A but for $m = 4$ and $m = 12$. Parameter values: $\varepsilon = 0.01$, $D = 0.001$, $\delta = (1 - \varepsilon)^3/2(m - 1)$, $\alpha = 1$, $M = 500$, $\theta = \theta^* - D$, $n=400$. }
\label{Fig7}
\end{figure}

Figure \ref{Fig7} illustrates the theorem numerically. First we assumed that the relevant set $\mathcal{Y}$ consists of $m =2$ items. One of them is considered as ``known'' to the neuron (Fig. \ref{Fig7}A, green). Its informational content, $\bfx_{M+1}$, satisfies {the condition} $\langle \bfw_0, \bfx_{M+1}\rangle>\theta$, i.e., this stimulus evokes membrane potential above the threshold {at $t=t_0$}. Consequently, the neuron detects this stimulus selectively as described in Sect. \ref{Section4.1}. For the second relevant stimulus (Fig. \ref{Fig7}A, orange), however, we have  $\langle \bfw_0,\bfx_{M+2} \rangle < \theta$. Therefore, the neuron cannot detect such a stimulus alone. The background stimuli from the set $\mathcal{S}(\mathcal{M})$ are  also sub-threshold (Fig. \ref{Fig7}A, back curves).

During the learning phase, the neuron receives $M=500$ background and $m=2$ relevant stimuli. The relevant stimuli from the set $\mathcal{S}(\mathcal{Y})$ appear simultaneously, i.e., they are temporarily associated. The synaptic efficiency changes during the learning phase by action of the relevant stimuli. Therefore, the membrane potential, $y(t) = \langle\bfw(t,{\bfw_0}),\bar{\bfs}(t) \rangle$, progressively increases when the relevant stimuli arrive (Fig. \ref{Fig7}A, green area). These neuronal adjustments give rise to a new functionality.

At some time instant (marked by red circle in Fig. \ref{Fig7}A) the neuron becomes responsive to the new relevant stimulus (Fig. \ref{Fig7}A, orange), which is synchronized with the ``known'' one. Note that all other background stimuli, that show no temporal associativity, remain below the threshold (Fig. \ref{Fig7}A, black traces). Thus, after a transient period, the neuron learns new stimulus. Once the learning is over, the neuron detects selectively  either of the two relevant stimuli.

The procedure just described {can be used to associate together} more than two relevant stimuli. Figure \ref{Fig7}B shows examples for $m=4$ and $m=12$. In both cases the neuron was able to learn all relevant stimuli, whilst rejecting all background ones. We observed, however, that  increasing the number of {uncorrelated} information items to be learnt, i.e. the value of $m$, {reduces the gap} between  firing thresholds and the membrane potentials evoked by background stimuli. In other words, the neuron does detect the assigned group of new stimuli, but with lower accuracy. This behavior is {consistent with the theoretical bound} on $\theta$ prescribed in the statement of Theorem \ref{theorem:new_memories}.

\section{Discussion}
\label{sec:discussion}

Theorems \ref{theorem:selectivity}--\ref{theorem:new_memories} and our numerical simulations demonstrate that the extreme neuronal selectivity to single and multiple stimuli, and the capability to learn uncorrelated stimuli observed in a range of empirical studies \cite{Quiroga:2005}, \cite{Quiroga:2009}, \cite{Quiroga:2015}  can be explained by simple functional mechanisms implemented in single neurons. The following {basic phenomenological properties have been used to arrive to this conclusion}: i) the dimensionality $n$ of the  information content and neurons is sufficiently large,  ii) a perceptron neuronal model, Eq. (\ref{eq:neuron_model}), is an adequate representation of the neuronal response to stimuli, and iii) plasticity of the synaptic efficiency is governed by Hebbian rule (\ref{eq:hebbian_oja}). A crucial consequence of our study is that no \textit{a priori} assumptions on the structural organization of neuronal ensembles are necessary for explaining basic concepts of static and dynamic memories.

Our approach does not take into account more advanced neuronal behaviors {reproduced} by, e.g., models of spike-timing dependent plasticity \cite{Markram1997} and firing threshold adaptation \cite{fontaine2014spike}. Nevertheless, our model captures essential properties of neuronal dynamics and as such is generic enough for the purpose of functional description of memories.

Firing threshold adaption, as reported in \cite{fontaine2014spike}, steers firing activity of a stimulated neuron to a homeostatic state. In this state, the value of the threshold is just large/small enough to maintain reasonable firing rate without over/under-excitation. In our model, such a mechanism {could be achieved} by {setting} the value of $\theta$ sufficiently close to {the highest feasible values specified} in Theorems \ref{theorem:selectivity} and \ref{theorem:selectivity_multiple}.

In addition to rather general model of neuronal behavior, another major theoretical assumption of our work was the presumption that stimuli informational content is drawn from an equidistribution in a unit ball $B_n(1)$. This assumption, however, can be relaxed, and results of Theorems \ref{theorem:selectivity}--\ref{theorem:new_memories} generalized to product measures. Key ingredients of such generalizations are provided in \cite{GorbanTyukin:NN:2017}, and their practical feasibility is illustrated  by numerical simulations with  information items randomly drawn from a hypercube (Figs. \ref{Fig5}--\ref{Fig7}).

Our theoretical and numerical analysis revealed an interesting hierarchy of cognitive functionality implementable at the level of single neurons. We {have shown} that cognitive functionality develops with the dimensionality or connectivity parameter $n$ of single neurons. This reveals explicit relationships between {levels} of the neural connectivity in living organisms and different cognitive behaviors such organisms can exhibit (cf. \cite{Lobov2017}). As we can see from Theorems \ref{theorem:selectivity}, \ref{theorem:selectivity_multiple} and Figs. \ref{Fig5}, \ref{Fig6}, the ability {to form} static memories increases monotonically with $n$. The increase of cognitive functionality, however, occurs in steps.

 {For $n$ small ($n\in[1,10]$),} neuronal selectivity to a single stimulus does not form. {It emerges} rapidly when the dimension parameter $n$ exceeds some critical value, around $n=10\div20$ (see Fig. \ref{Fig5}A). This constitutes the first critical transition. Single neurons become selective to single information items. The second critical transition occurs at significantly larger dimensions, around $n=100-400$ (see Fig. \ref{Fig6}). At this second stage the neuronal selectivity to multiple {\it uncorrelated} stimuli develops. The ability to respond selectively to a given set of multiple uncorrelated information items is apparently crucial for rapid learning ``by temporal association'' in such neuronal systems. This learning ability as well as formation of dynamic memories are justified by Theorem \ref{theorem:new_memories} and illustrated in Fig. \ref{Fig7}.

In the core of our mathematical arguments are the concentration of measure phenomena exemplified in \cite{GorTyu:2016,GorbanTyukin:RSTA:2017} and  stochastic separation theorems \cite{GorbanTyukin:NN:2017,GorTyuRom2016}. Some of these results, which have been central in the proofs of Theorem \ref{theorem:selectivity_multiple} and \ref{theorem:new_memories}, namely, the statements that random i.i.d. vectors from equidistributions in $B_n(1)$ and product measures are almost orthogonal with probability close to one, are tightly related to the notion of effective dimensionality of spaces based on  $\epsilon$-{\it quasi\-or\-tho\-go\-na\-lity} introduced in \cite{Hecht,Kurkova}. In these works the authors demonstrated that in high dimensions there exist exponentially large sets of quasiorthogonal vectors.  In \cite{GorTyu:2016}, however, as well as in our current work (see Lemma \ref{lem:almost_orthogonal}) we demonstrated that not only such sets exist, but also that they are typical.

Finally, we note that the number of multiple stimuli that can be selectively detected by single neurons is not extraordinarily large. In fact, as we have shown in   Figs. \ref{Fig6} and \ref{Fig7}, memorizing $8$ information items  at the level of single neurons requires more than $400$ connections. This suggests, that not only new memories are naturally packed {\it in quanta}, but also that there is a limit on this number that is associated with the cost of implementation of such a functionality. This cost is the number of individual functional synapses. Balancing the costs in living beings is of course a subject of selection and evolution. Nevertheless, as our study have shown, there is a clear functional gain that these costs may be paid for.

\section{Conclusion}
\label{sec:conclusion}

In this work we analyzed the striking consequences of the abundance of signalling routes for functionality of neural systems. We demonstrated that complex cognitive functionality derived from extreme selectivity to external stimuli and rapid learning of new memories at the level of single neurons can be explained by the presence of multiple signalling {routes} and simple physiological mechanisms. At the basic level, these mechanisms can be reduced to a mere perceptron-like behavior of neurons in response to stimulation and a Hebbian-type learning governing changes of the synaptic efficiency.

The observed phenomenon is robust. Remarkably, a simple generic model  offers a clearcut mathematical explanation of a wealth of empirical evidence related to \textit{in-vivo} recordings of ``Grandmother'' cells, ``concept'' cells, and rapid learning at the level of individual neurons \cite{Quiroga:2005,Quiroga:2009,Quiroga:2015}. {The results can also shed light on the question why Hebbian learning may give rise to neuronal selectivity in prefrontal cortex \cite{lindsay2017hebbian} and explain why adding single neurons to deep layers of artificial neural networks is an efficient way to acquire novel information while preserving previously trained data representations  \cite{draelos2016neurogenesis}.}

Finding simple laws explaining complex behaviours has always been the driver of progress in Mathematical Biology and Neuroscience. Numerous examples of such simple laws can be found in the literature (see e.g. \cite{roberts2014can,jurica2013sensory,gorban2016evolution,perlovsky2006toward}). Our results not only provide a {simple} explanation of the reported empirical evidence but also {suggest} that such a behavior {might be} inherent to {neuronal systems and hence} organisms that {operate} with high-dimensional informational content. In such systems, complex cognitive functionality at the level of elementary units, i.e., single neurons, occurs naturally.  The higher the dimensionality, the stronger the effect. In particular, we have shown that the memory capacity in ensembles of single neurons grows exponentially with the neuronal dimension. Therefore, from the evolutionary point of view, accommodating large number of signalling routes converging onto single neurons is advantageous despite the increased metabolic costs.

The considered class of neuronal models, being generic, is of course a simplification. It does not capture spontaneous firing, signal propagation in dendritic trees, and many other physiologically relevant features of real neurons. Moreover, in our theoretical assessments we assumed that the informational content processed by neurons is sampled from an equidistribution in a unit ball. The results, however, can already be generalized to product measure distributions (see, e.g., \cite{GorbanTyukin:NN:2017}).  Generalizing the findings to models offering better physiological realism is the focus of our future works.


\begin{acknowledgements}
This work has been supported by Innovate UK grants KTP009890 and KTP010522, by the Spanish Ministry of Economy and Competitiveness under grant FIS2014-57090-P, the Russian Federation Ministry of Education state assignment (No. 8.2080.2017/4.6), ``Initiative scientific project'' of the main part of the state plan of the Ministry of Education and Science of Russian Federation (task No. 2.6553.2017/BCH Basic Part), and by the Russian Science Foundation  project 15-12-10018 (numerical assessment and results). Alexander N. Gorban was supported by  the Ministry of Education and Science of Russian Federation (Project No. 14.Y26.31.0022).
\end{acknowledgements}



\appendix
\section{Dynamics of coupling weights}
\label{AppendixDynamicsW}

The following results demonstrate that the neuronal model provided in Section \ref{sec:model_neuron} is well-posed.

\begin{lemma}\label{lem:learning_rule} Consider  (\ref{eq:neuron_model}), (\ref{eq:hebbian_oja}) {with the function $\bfs(\cdot,\bfx)$, $\bfx\in\Real^n$ defined as in (\ref{eq:stimulus_definition}).}  Then

\begin{enumerate}
\item[1)] solutions $\bfw(\cdot,\bfw_0)$ of (\ref{eq:hebbian_oja}) are defined for all $t\geq t_0$, {and are} unique and bounded in forward time.
\end{enumerate}
If, in addition, {$\theta\geq 0$ and} there exist numbers $L,\delta>0$ such that:
\begin{equation}\label{eq:persistent_excitation}
\int_{t}^{t+L} {v(\bfs(\tau,\bfx),\bfw(\tau,\bfw_0),\theta) \langle \bfs(\tau,\bfx),\bfw(\tau,\bfw_0) \rangle^2} \, \mathrm{d}\tau > \delta, \ \ \  \forall t\geq t_0,
\end{equation}
then

\begin{enumerate}
\item[2) ] $\bfx/\|\bfx\|$ is an attractor, that is:
\begin{equation}\label{eq:weights_asymptote}
\lim_{t\rightarrow \infty} \bfw(t,\bfw_0)=\frac{\bfx}{\|\bfx\|}.
\end{equation}
\end{enumerate}
\end{lemma}

\vspace{0.2cm}

\noindent
{\it Proof of Lemma \ref{lem:learning_rule}}.

\noindent
1. The right-hand side of (\ref{eq:hebbian_oja}) is continuous in $\bfw$ and piece-wise continuous in $t$ with finite number of discontinuities of the first kind in any finite interval containing $t_0$, independently on the values of $\bfw$. Hence, in accordance with Peano Theorem, solutions of (\ref{eq:hebbian_oja}) are defined on some non-empty interval containing $t_0$. Let $\mathcal{T}$ be the maximal interval of this solution's definition (to the right of $t_0$). {Since the right-hand side of (\ref{eq:hebbian_oja}) is locally Lipschitz in $\bfw$ the solution $\bfw(\cdot,\bfw_0)$ is uniquely defined on $\mathcal{T}$.}

{To show that $\mathcal{T}=[t_0,\infty)$} consider
\[
V(\bfw)=1-\|\bfw\|^2.
\]
In the interval $\mathcal{T}$ we have:
\[
\dot{V}= -2 \alpha  v y^2 V.
\]
Given that $vy^2 \geq 0$, the above expression implies that
\[
|1 - \|\bfw_0\|^2| \geq |1-\|\bfw(t,\bfw_0)\|^2|\geq \|\bfw(t,\bfw_0)\|^2 - 1.
\]
Consequently,
\begin{equation}\label{eq:weights_bound}
\|\bfw(t,\bfw_0)\| \le \left(1 + |1 - \|\bfw_0\|^2|\right)^{\frac{1}{2}}
\end{equation}
{for all $t\geq t_0$, $t\in \mathcal{T}$. Let $t_1$ be an arbitrary point in the interval $\mathcal{T}$. Recall that the right-hand side of (\ref{eq:hebbian_oja}) is continuous and locally Lipschitz with respect to $\bfw$ (uniformly in $t$). Thus (\ref{eq:weights_bound}) implies existence of some $\Delta(\bfw_0,\bfx)>0$, independent on $t_1$, such that the solution $\bfw(\cdot,\bfw_0)$ is defined on the interval $[t_0,t_1+\Delta(\bfw_0,\bfx)]$. Given that $t_1$ was chosen arbitrarily in $\mathcal{T}$, we can conclude that $\mathcal{T}=[t_0,\infty)$ (cf. Theorem 3.3 \cite{Khalil_2002}).}

\noindent
2. For the sake of convenience, we denote
\[
p(t)= v(\bfs(t,\bfx),\bfw(t,\bfw_0),\theta) \langle \bfs(t,\bfx),\bfw(t,\bfw_0) \rangle^2.
\]
Condition (\ref{eq:persistent_excitation}) assures that both $\bfx\neq0$, $\bfw_0\neq 0$. Moreover, since $V(\bfw(t,\bfw_0))$ is defined for all $t\geq t_0$, we can conclude that
\[
{|V(t)| = \left|V_0 e^{-{2}\alpha \int_{t_0}^{t} p(\tau) \mathrm{d}\tau}\right|  \leq |V_0| e^{-{2}\alpha \delta \left\lfloor\frac{t-t_0}{L}\right\rfloor}.}
\]
Hence
\begin{equation}\label{eq:weights_normalized}
\lim_{t\rightarrow \infty} \|\bfw(t,\bfw_0)\|=1.
\end{equation}
Consider:
\[
\begin{array}{ll}
\bfw(t,\bfw_0)=& e^{-\alpha \int_{t_0}^{t} p(\tau) \mathrm{d}\tau } \bfw_0 +\\
& {\alpha \left[\int_{t_0}^{t} e^{-\alpha \int_{\tau}^{t} p(s) \mathrm{d}s } v(\bfs(\tau,\bfx),\bfw(\tau,\bfw_0),\theta) \langle \bfs(\tau,\bfx),\bfw(\tau,\bfw_0) \rangle\sum_j c(\tau-\tau_j)\,  \mathrm{d}\tau\right]} \bfx.
\end{array}
\]
Observe that the first term decays exponentially to $0$, whereas the second term is proportional to $\bfx$. {Moreover, since $\theta\geq 0$, the  term
$v(\bfs(\tau,\bfx),\bfw(\tau,\bfw_0),\theta)\langle \bfs(\tau,\bfx),\bfw(\tau,\bfw_0) \rangle \geq 0$ for all $\tau\geq t_0$. Hence the coefficient in front of $\bfx$ is non-negative.} This, combined with (\ref{eq:weights_normalized}), {implies that (\ref{eq:weights_asymptote}) holds}. $\square$

\vspace{0.2cm}

{Note that Lemma \ref{lem:learning_rule} apply to stimuli classes that are broader than the one defined by (\ref{eq:stimulus_definition}), (\ref{eq:spike_shape_definition}). The results hold e.g. for the functions $c(\cdot)$ in (\ref{eq:stimulus_definition}) that are non-negative, piece-wise continuous, and bounded. On the other hand, to determine convergence and asymptotic properties of $\bfw(\cdot,\bfw_0)$  for $t\geq t_0$ (part 2 of the lemma) one needs to check that condition (\ref{eq:persistent_excitation}) holds. A drawback of this condition is that it requires availability of signals  $v(\bfs(t,\bfx),\bfw(t,\bfw_0),\theta)$,  $\langle \bfs(t,\bfx),\bfw(t,\bfw_0) \rangle$ for all $t\geq t_0$.}

{For  $c(\cdot)$ specified by (\ref{eq:stimulus_definition}) this latter condition can be drastically simplified. To see this, let us}  get a somewhat deeper geometrical insight {into} the dynamics of $\bfw$ governed by (\ref{eq:hebbian_oja}). {In order to bring the discussion in line with the question of neuronal selectivity, consider the stimuli sets (\ref{eq:M_set}), (\ref{eq:stimuli_M}) with $\mathcal{Y}=\{\bfx_{M+1}\}$, and suppose that stimuli $\bfs(\cdot,\bfx_i)$, $i=1,\ldots,M$ do not evoke any neuronal responses, i.e., $v(\bfs(\cdot,\bfx_i),\bfw(\cdot,\bfw_0),\theta) = 0$ for all $i=1,\ldots,M$. Hence no changes in $\bfw$ occur if the stimulus $\bfs$ in (\ref{eq:hebbian_oja}) is any of $\bfs(\cdot,\bfx_i)$, $i=1,\ldots,M$.}

 {Consider system (\ref{eq:hebbian_oja}) with $\bfs(\cdot,\bfx_{M+1})$}. {The variable $\bfw$ may change only over those intervals of $t$ when $\bfs(\cdot,\bfx_{M+1})\neq 0$. Between these intervals $\bfw(t,\bfw_0)$ is constant. Let the stimulus be persistent in the sense that for any $t'\geq t_0$ there is a $t''$ such that $\bfs(t'',\bfx_{M+1})\neq 0$.} Thus, without {loss} of generality and for the purposes of assessing asymptotic behavior of $\bfw(t,\bfw_0)$ at $t\rightarrow\infty$ {variable $\bfs(t,\bfx_{M+1})$ in (\ref{MembranePotential}) -- (\ref{eq:hebbian_oja}) may be replaced  with $\bfx_{M+1}$.}

{Recall that $\bfw(t,\bfw_0)$ can be represented as a sum
\[
\bfw(t,\bfw_0)={w^{\ast}}(t,\bfw_0) \bfw^\ast + \bfw^{\bot}(t,\bfw_0), \ {w^{\ast}}(t,\bfw_0)=\langle\bfw(t,\bfw_0),\bfw^\ast\rangle,
\]
where $\bfw^\ast$ is defined in (\ref{eq:OrthonormalBasis}) and $\bfw^{\bot}\in L^{\bot}$. In this representation,
\[
\begin{split}
&\dot{\bfw}= \dot{w}^{\ast} \bfw^\ast + \dot{\bfw}^{\bot}= \alpha f(\langle \bfx_{M+1}, {w^{\ast}} \bfw^\ast + \bfw^{\bot} \rangle - \theta) \langle \bfx_{M+1}, {w^{\ast}} \bfw^\ast + \bfw^{\bot} \rangle (\bfx_{M+1} -\\
 &\langle \bfx_{M+1}, {w^{\ast}} \bfw^\ast + \bfw^{\bot} \rangle [{w^{\ast}} \bfw^\ast + \bfw^{\bot}] ) = \left[\alpha f({w^{\ast}}\|\bfx_{M+1}\|-\theta)\|\bfx_{M+1}\|^2(1 - {w^{\ast}}^2)\right] {w^{\ast}}\bfw^\ast  -\\
  &\left[\alpha f ({w^{\ast}}\|\bfx_{M+1}\|-\theta) \|\bfx_{M+1}\|^2 {w^{\ast}}^2\right] \bfw^{\bot}
\end{split}
\]
or, equivalently,
\begin{eqnarray}
\dot{w}^\ast&=& \alpha \|\bfx_{M+1}\|^2 f({w^{\ast}}\|\bfx_{M+1}\|-\theta)(1 - {w^{\ast}}^2) {w^{\ast}} \label{eq:App_Case1}\\
\dot{\bfw}^{\bot}&=&- \left[\alpha f ({w^{\ast}}\|\bfx_{M+1}\|-\theta) \|\bfx_{M+1}\|^2 {w^{\ast}}^2\right] \bfw^{\bot} \label{eq:App_Case2}.
\end{eqnarray}
Obviously, $L^{\|}$, $L^{\bot}$, and the set
\[
\mathcal{W}(\bfx_{M+1},\theta)=\{({w^{\ast}},\bfw^{\bot}), \ {w^{\ast}}\in\Real, \bfw^{\bot}\in L^{\bot}\ | {w^{\ast}} \|\bfx_{M+1}\| - \theta \leq 0 \}
\]
are invariant with respect to (\ref{eq:hebbian_oja}). Let $\bfx_{M+1}\neq 0$, $\theta\geq 0$, and $\bfw_0\notin \mathcal{W}(\bfx_{M+1},\theta)$. Then two non-trivial alternatives  (Fig. \ref{FigA1}) are possible}:

\begin{figure}[!h]
\centering{\epsfig{file = 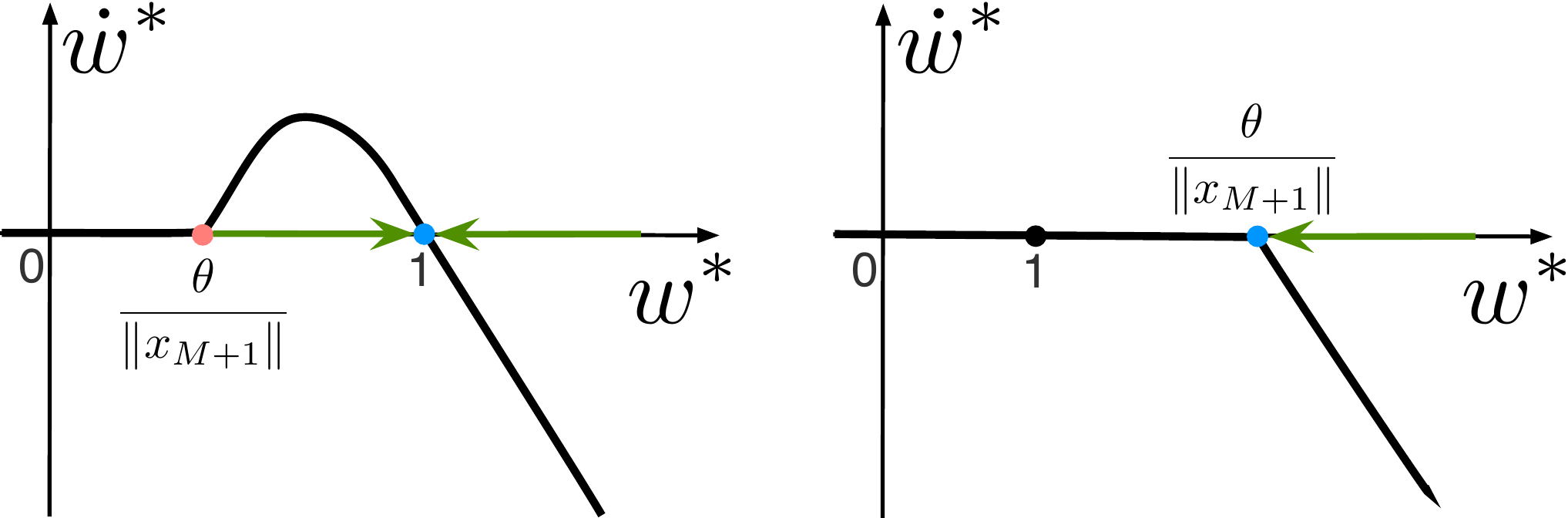, width = 0.7\textwidth}}
\caption{Sketch of the dynamics of {${w^{\ast}}$}. Thick black curve {shows the r.h.s. of (\ref{eq:App_Case1}) as a function of ${w^{\ast}}$} for two cases: $\theta < \|x_{M+1}\|$ (left) and $\theta > \|x_{M+1}\|$ (right). Blue (red) dots correspond to stable (unstable) equilibria. Green arrows mark trajectories. In the first case (left) ${w^{\ast}}$ tends to 1, whereas in the second (right) it goes asymptotically to $\theta / \|x_{M+1}\|$.}
\label{FigA1}
\end{figure}

\begin{itemize}
\item[{\bf A:}] If $\theta < \|\bfx_{M+1}\|$ then {${w^{\ast}}(t,\bfw_0) \to 1$ and, according to (\ref{eq:App_Case2}), $\bfw^{\bot}(t,\bfw_0)\to 0$ as} $t\to \infty$. Thus,
\[
\lim_{t\to \infty} \bfw(t) = \frac{\bfx_{M+1}}{\|\bfx_{M+1}\|} = \bfw^\ast.
\]

\item[{\bf B:}] If $\theta \ge \|\bfx_{M+1}\|$ then ${w^{\ast}}(t,\bfw_0) \to \theta/\|\bfx_{M+1}\|$ as $t\to \infty$. {There is no guarantee, however that $\bfw^{\bot}(t,\bfw_0)$ converges to the origin asymptotically.} Thus, {there is a $\bfw^{\bot}_{\infty}\in L^\bot$}:
\[
\lim_{t\to \infty} \bfw(t) = \frac{\theta}{\|\bfx_{M+1}\|}\bfw^\ast + {\bfw^{\bot}_{\infty}}.
\]
\end{itemize}
{The above result can now be formalized as
\begin{lemma}\label{lem:learning_rule:2} Consider  (\ref{eq:neuron_model}), (\ref{eq:hebbian_oja}) with the function $\bfs(\cdot,\bfx)$, $\bfx\in\Real^n$ defined as in (\ref{eq:stimulus_definition}). Let $\theta\geq 0$ and $\langle\bfw_0,\bfx \rangle>\theta$. Furthermore, let the stimulus $\bfs(\cdot,\bfx)$ be persistent in the sense that for any $t'\geq t_0$ there is a $t''>t'$ such that $\bfs(t'',\bfx)\neq 0$. Then the following alternatives hold:
\begin{itemize}
\item[1)] If $\theta < \|\bfx\|$ then $\lim_{t\rightarrow\infty}\bfw(t,\bfw_0)=\bfx/\|\bfx\|$.
\item[2)] If $\theta \geq \|\bfx\|$ then $\lim_{t\rightarrow\infty}\langle\bfw(t,\bfw_0),\bfx/\|\bfx\|\rangle=\theta/\|\bfx\|$.
\end{itemize}
\end{lemma}
}

{Note that alternative 1) in Lemma \ref{lem:learning_rule:2} is equivalent to the second statement of Lemma \ref{lem:learning_rule}. Alternative 2) corresponds to the case when condition (\ref{eq:persistent_excitation}) of Lemma \ref{lem:learning_rule} is not satisfied.}


\section{Proof of Theorem \ref{theorem:selectivity}}
\label{ProofTheorem1}

1. Let us first assume that $\|\bfw\|=1$. Notice that the condition
\begin{equation}\label{eq:equivalent_functional}
\langle \bfw, \bfx_i \rangle \le \theta   \ \ \ \forall  \bfx_i\in\mathcal{M},
\end{equation}
assures that $v=0$ and hence $\bfs_i \in \mathrm{Silent}(\mathcal{S}(\mathcal{M}),(\bfw,\theta)) \ \ \forall \bfs_i\in\mathcal{S}(\mathcal{M})$.

In this case the neuron is silent for all stimuli except $\bfs_{M+1}$ that does evoke a response by construction. Therefore, it is sufficient to estimate the probability that (\ref{eq:equivalent_functional}) holds.

Let $\mathcal{C}_n(\bfw,\theta)$ be the spherical cap:
\[
\mathcal{C}_n(\bfw,\theta)=\{ \bfx\in B_n(1) \ | \ \langle \bfw, \bfx \rangle > \theta  \}.
\]
Then the ratio of volumes $\mathcal{V}(\mathcal{C}_n(\bfw,\theta))/\mathcal{V}(B_n(1))$ is the probability that a random vector $\bfx_i \in \mathcal{C}_n(\bfw,\theta)$.  Observe that
\[
\frac{\mathcal{V}(\mathcal{C}_n(\bfw,\theta))}{\mathcal{V}(B_n(1))}\leq \frac{1}{2}(1-\theta^2)^{\frac{n}{2}}.
\]
Thus, the probability that all $\bfx_i\in\mathcal{M}$ are outside  the cap $\mathcal{C}_n(\bfw,\theta)$ is bounded from below:
\begin{equation}
\label{P_preliminar}
P = \left [1 -  \frac{\mathcal{V}(\mathcal{C}_n(\bfw,\theta))}{\mathcal{V}(B_n(1))} \right ]^M \ge \left[1- \frac{1}{2}(1-\theta^2)^{\frac{n}{2}}\right]^M,
\end{equation}
which is equivalent to (\ref{eq:selectivity_1}), given that $\|\bfw\|=1$.

Let $\|\bfw\|\neq 1$. {Noticing that, for $\|\bfw\|> 0$
\[
\langle \bfw, \bfx_i \rangle \le \theta   \ \ \ \forall  \bfx_i\in\mathcal{M} \ \Leftrightarrow \ \langle \bfw/\|\bfw\|, \bfx_i \rangle \le \theta/\|\bfw\| \ \ \ \forall  \bfx_i\in\mathcal{M},
\]
and substituting $\theta/\|\bfw\|$ in place of $\theta$ in  (\ref{P_preliminar}) results in  (\ref{eq:selectivity_1}).}

\vspace{0.2cm}

\noindent
2. Let us show that for $(\bfw,\theta)\in\Omega_D$ the neuron detects the relevant stimulus $\bfs_{M+1}$, i.e., $v>0$. Using (\ref{OM_D}) we observe that
\[
\begin{split}
\langle  \bfw, \bfx_{M+1} \rangle - \theta &= \langle  \bfw - \bfw^{\ast}, \bfx_{M+1} \rangle + \|x_{M+1}\| - \theta \ge \langle  \bfw - \bfw^{\ast}, \bfx_{M+1} \rangle + D \ge\\
&\ge -\|\bfw - \bfw^{\ast}\|\|\bfx_{M+1}\| + D > D(1 - \|\bfx_{M+1}\|)\geq 0,
\end{split}
\]
implying that $\bfs_{M+1}\in\mathrm{Activated}(\mathcal{Y},(\bfw,\theta))$.

Let us evaluate the probability that the neuron rejects all background stimuli for all $(\bfw,\theta)\in\Omega_D$. According to (\ref{OM_D}) the following holds:
\[
\frac{\theta}{\|\bfw\|} \geq \frac{\|\bfx_{M+1}\| - 2 D}{1 + D}, \ \ \ \forall  (\bfw,\theta)\in\Omega_D.
\]
Moreover, $\|\bfx_{M+1}\| \geq 1-\varepsilon$ with probability $p=1-(1-\varepsilon)^n$. Therefore,  with probability larger or equal to $p$, the ratio $\frac{\theta}{\|\bfw\|}$ us bounded from below as:
\[
\frac{\theta}{\|\bfw\|} \geq  \frac{1-\varepsilon - 2 D}{1 + D}.
\]
Finally, since the value of $\varepsilon$ can be chosen arbitrarily in the interval  $(0,1-2D)$ and taking into account that the right-hand side of (\ref{P_preliminar}) is a monotone and increasing function with respect to $\theta$ in the interval $[0,1]$,  estimate (\ref{eq:selectivity_2}) immediately follows from  (\ref{P_preliminar}) and (\ref{eq:selectivity_1}).  $\square$

\section{Proof of Corollary \ref{LEMMA_1}}
\label{ProofLEMMA_1}

Consider (\ref{eq:selectivity_1}) and denote
\begin{equation}
\label{po}
z = \frac{1}{2}
\left[1 - \frac{\theta^2}{\|\bfw\|^2}
\right]^\frac{n}{2}, \ \ \
\phi = (1-z)^{\overline{M}}.
\end{equation}
According to (\ref{po}), $(1-z)^M\geq \phi$ for all $0<M\leq \overline{M}$. Given that $z\in (0,1)$, from Eq. (\ref{po}) we get $\ln(\phi)= \overline{M} \ln(1-z)$. Recall that $\ln(1-z) > -z/(1-z)$, $\forall  z\in (0,1)$. Thus, we can conclude that
\begin{equation}
\label{APPC_1}
\overline{M} > -\ln(\phi)  \frac{1-z}{z} =-\ln(\phi)(z^{-1} -1)= -\ln(\phi)\left(2 e^{\alpha n}- 1\right),
\end{equation}
where $\alpha$ is given by (\ref{LEM:1}). Thus, according to (\ref{APPC_1}), for $0<M\leq -\ln(\phi)\left(2 e^{an}- 1\right) < \overline{M}$ the following holds
\[
P( \bfs_i \in \mathrm{Silent}(\mathcal{S}(\mathcal{M}),(\bfw,\theta)) \ \forall  \bfs_i\in\mathcal{S}(\mathcal{M}) \big| \ \bfw,\theta) \ge \phi.
\]
$\square$

\section{Proof of Theorem \ref{theorem:selectivity_multiple}} \label{ProofTheorem2}

The proof of the Theorem is essentially contained in Lemmas \ref{lem:almost_orthogonal} and \ref{lem:k-tuples:ball:correlated} (Sect. \ref{AppendixAux}). Consider the set $\mathcal{Y}$. With probability $(1-(1-\varepsilon)^n)^m$, all elements $\bfx_i\in\mathcal{Y}$ satisfy the condition $\|\bfx_i\|\geq 1-\varepsilon$. Hence, using Lemma \ref{lem:almost_orthogonal} we have that the following inequality
\[
|\langle \bfx_i, \bfx_j \rangle| \leq \frac{\delta}{1-\varepsilon}, \ \ \forall \bfx_i,\bfx_j\in\mathcal{Y}, \ i\neq j
\]
holds with probability
\[
p_0\geq  (1-(1-\varepsilon)^n)^m \prod_{d=1}^{m-1} \left(1-d(1-\delta^2)^{\frac{n}{2}}\right).
\]
This implies that, with probability $p_0$, the following conditions are met
\[
\|\bfx_i\|\geq 1-\varepsilon, \ \ \  -\frac{(m-1)\delta}{1-\varepsilon}\leq \sum_{j=1, \ j\neq i}^m \langle \bfx_i,\bfx_j \rangle \leq \frac{(m-1)\delta}{1-\varepsilon},  \ \ \forall \ \bfx_i\in\mathcal{Y}.
\]
Consider $\ell(\bfx)=\langle \bfw^{\ast}, \bfx \rangle - \theta^\ast+D$. Invoking Lemma \ref{lem:k-tuples:ball:correlated} and setting $\beta_1=\delta/(1-\varepsilon)$, $\beta_2=-\delta/(1-\varepsilon)$, we can conclude that,  with probability $p_0$,
\[
\ell(\bfx)\geq D, \ \ \forall  \bfx\in\mathcal{Y}.
\]
In fact, we can conclude {that with probability $p_0$}
\[
\ell_0(\bfx)=\langle \bfw, \bfx \rangle - \theta = \ell(\bfx) + \langle \bfw-\bfw^{\ast},\bfx\rangle - \theta+(\theta^\ast-D) > 0, \ \ \forall \  (\bfw,\theta)\in\Omega_D, \ \bfx\in\mathcal{Y}.
\]
Thus, the probability that $\ell_0(\bfx)>0$ for all $\bfx\in\mathcal{Y}$ and that $\ell_0(\bfx)\leq 0$ for all $\bfx\in\mathcal{M}$ is bounded from below by
\[
(1-(1-\varepsilon)^n)^m \prod_{d=1}^{m-1} \left(1-d(1-\delta^2)^{\frac{n}{2}}\right) \left[1-\frac{1}{2}\left(1-\frac{\theta^2}{\|\bfw\|^2}\right)^{\frac{n}{2}}\right]^{M}.
\]
Noticing that $\|\bfw\|\leq 1+D$, we can conclude that (\ref{eq:prob_selectivity2}) holds. $\square$

\section{Proof of Theorem \ref{theorem:new_memories}}\label{ProofTheorem3}

According to Lemma \ref{lem:learning_rule}, solutions $\bfw(t,\bfw_0)$ are defined for all $t\geq t_0$. Moreover, condition 1 of the theorem and Lemma \ref{lem:learning_rule} imply that
\begin{equation}\label{eq:w_limit}
\lim_{t\rightarrow \infty} \bfw(t,{\bfw_0})=\frac{\sum_{i=1}^m \bfx_{M+i}}{\|\sum_{i=1}^m \bfx_{M+i}\|}=\bar{\bfx}/\|\bar{\bfx}\|=\bfw^\ast.
\end{equation}
Let $D>0$ be chosen so that
{\[
0<\theta + D \leq \theta^\ast.
\]
Given that $0<\theta<\theta^\ast$,} such $D$s always exist. Equation (\ref{eq:w_limit}) implies that there is a $t_1(D)> t_0$ such that
\begin{equation}\label{eq:w_limit:2}
{\|\bfw(t,{\bfw_0})-\bfw^\ast\|<D,  \ \theta\in(0,\theta^\ast-D] \  \forall  \ t\geq t_1(D).}
\end{equation}
{The theorem now} follows immediately from Theorem \ref{theorem:selectivity_multiple}. $\square$

\section{Auxiliary results}\label{AppendixAux}
\begin{lemma}[cf.  Gorban et. al. 2016, \cite{GorTyu:2016}]\label{lem:almost_orthogonal} Let $\mathcal{Y}=\{\bfx_1,\bfx_2,\dots,\bfx_k\}$ be a set of $k$ i.i.d. random vectors  from the equidistribution in the unit ball $B_n(1)$. Let $\delta,r\in(0,1)$, and suppose that $\|\bfx_i\|\geq r$, for all $i\in\{1,\dots,k\}$.

Then the probability that the elements of $\mathcal{Y}$ are pair-wise $\delta/r$-orthogonal, that is
\[
|\cos (\angle (\bfx_i,\bfx_j) )| \leq \frac{\delta}{r} \ \mbox{for all} \ i\neq j \  \ i,j\in\{1,\dots,k\},
\]
is bonded from below as
\[
\begin{split}
&\mathcal{P}\left(|\cos (\angle (\bfx_i,\bfx_j) )| \leq \frac{\delta}{r} \ \forall \ i,j\in\{1,\dots,k\}, \ i\neq j \ \left| \ \|\bfx_i\|\geq r, \  1\leq i\leq k \right.\right) \\
&\ \ \ \ \ \ \ \ \ \ \ \ \ \ \ \ \ \ \geq  \prod_{d=1}^{{k}-1} \left(1-d \left(1 -\delta^2\right)^{\frac{n}{2}} \right).
\end{split}
\]
\end{lemma}
{\it Proof of Lemma \ref{lem:almost_orthogonal}}. Let $\bfx_i$, $i=1,\dots,k$ be random vectors satisfying conditions of the lemma.  Let $E_\delta(\bfx_i)$ be the delta-thickening of the largest equator of $B_n(1)$ that is orthogonal to $\bfx_i$. There is only one such equator, and it is uniquely determined by $\bfx_i$.
Consider the following probabilities:
\[
\begin{split}
&P(\bfx_2\in\ E_\delta(\bfx_1))\\
&P([\bfx_3\in\ E_\delta(\bfx_2)]\& [\bfx_3\in\ E_\delta(\bfx_1)])\\
&P([\bfx_4\in\ E_\delta(\bfx_3)]\& [\bfx_4\in\ E_\delta(\bfx_2)]\& [\bfx_4\in\ E_\delta(\bfx_1)])\\
& \cdots \\
&P([\bfx_k\in\ E_\delta(\bfx_{k-1})]\& \cdots \& [\bfx_k\in\ E_\delta(\bfx_1)]).
\end{split}
\]
Pick $\bfx_i,\bfx_j\in\mathcal{Y}$, $i\neq j$. Recall that, for any random events $A_1,\dots,A_k$, the probability
\begin{equation}\label{eq:probability_and_formula}
P(A_1 \& A_2 \& \cdots \& A_k) \geq 1 - \sum_{i=1}^k (1-P(A_i)).
\end{equation}
According to (\ref{eq:probability_and_formula}), the probability that $\bfx_i\in E_\delta(\bfx_j)$ is bounded from below by $1-\left(1 -\delta^2\right)^{\frac{n}{2}}$ (cf. \cite{GorTyu:2016}, Proposition 3; see also Fig. 1 in \cite{GorTyu:2016} for illustration). Then
\begin{equation}\label{eq:prob_orthogonal_events}
\begin{split}
&P(\bfx_2\in\ E_\delta(\bfx_1))\geq 1-\left(1 -\delta^2\right)^{\frac{n}{2}}\\
&P([\bfx_3\in\ E_\delta(\bfx_2)]\& [\bfx_3\in\ E_\delta(\bfx_1)])\geq 1-2\left(1 -\delta^2\right)^{\frac{n}{2}}\\
&P([\bfx_4\in\ E_\delta(\bfx_3)]\& [\bfx_4\in\ E_\delta(\bfx_2)]\& [\bfx_4\in\ E_\delta(\bfx_1)])\geq 1-3\left(1 -\delta^2\right)^{\frac{n}{2}}\\
& \cdots \\
&P([\bfx_k\in\ E_\delta(\bfx_{k-1})]\& \cdots \& [\bfx_k\in\ E_\delta(\bfx_1)]) \geq 1-(k-1)\left(1 -\delta^2\right)^{\frac{n}{2}}.
\end{split}
\end{equation}
The fact that $\bfx_i\in E_\delta(\bfx_j)$ combined with the condition that $\|\bfx_i\|\geq r$, $\|\bfx_j\|\geq r$ imply:
\[
|\cos(\angle (\bfx_i,\bfx_j) )| \leq \frac{\delta}{r}.
\]
Finally, given that $\bfx_1,\dots,\bfx_k$ are drawn independently {and that the distribution is rotationally invariant},  the probability that all vectors in $\mathcal{Y}$ are pair-wise orthogonal is the product of all probabilities in the left-hand side of (\ref{eq:prob_orthogonal_events}). Thus the statement follows. $\square$

\begin{lemma}\label{lem:k-tuples:ball:correlated} Let $\mathcal{Y}=\{\boldsymbol{x}_{1},\dots,\boldsymbol{x}_{m}\}$  be  a finite set from  $B_n(1)$. Let $\|\bfx_i\|\geq 1-\varepsilon$, $\varepsilon\in(0,1)$ for all $\bfx_i\in\mathcal{Y}$, and $\beta_1,\beta_2\in\Real$ be such that the following condition holds:
\begin{equation}\label{eq:k-tuples:assumption}
\beta_2 (m-1) \leq \sum_{j\in\{1,\dots,m\}, \ j\neq i} \langle \boldsymbol{x}_{i}, \boldsymbol{x}_{j}\rangle \leq \beta_1 (m-1) \ \mbox{for all} \ i=1,\dots,m.
\end{equation}
Consider 
\[
\begin{split}
\ell (\bfx) = \left\langle \frac{\bar{\boldsymbol{y}}}{\|\bar{\boldsymbol{y}}\|}, \bfx \right\rangle - \frac{1}{\sqrt{m}}\left(\frac{(1-\varepsilon)^2 + \beta_2 (m-1)}{\sqrt{1+(m-1)\beta_1}}\right), \ \bar{\boldsymbol{y}}=&\frac{1}{m}\sum_{i=1}^{m} \boldsymbol{x}_{i},
\end{split}
\]
and suppose that parameters $\beta_1,\beta_2$ satisfy:
\[
(1-\varepsilon)^2 + \beta_2 (m-1) > 0, \ 1+(m-1)\beta_1 > 0.
\]
Then
\begin{equation}\label{eq:k-tuple:functional}
\ell(\bfx_i) \geq  0 \ \mbox{for all} \ \bfx_i\in\mathcal{Y}.
\end{equation}
%
%
\end{lemma}
{\it Proof of Lemma \ref{lem:k-tuples:ball:correlated}}. Consider the set $\mathcal{Y}$. According to the lemma assumptions, $\|\boldsymbol{x}_{i}\|\geq 1-\varepsilon$ for some given $\varepsilon\in(0,1)$ and all $i=1,\dots,m$. Consider now the mean vector  $\bar{\boldsymbol{y}}$
\[
\bar{\boldsymbol{y}}=\frac{1}{m}\sum_{i=1}^{m} \boldsymbol{x}_{i},
\]
and evaluate the following inner products
\[
\left\langle \frac{\bar{\boldsymbol{y}}}{\|\bar{\boldsymbol{y}}\|}, \boldsymbol{x}_{i} \right\rangle=\frac{1}{m \|\bar{\boldsymbol{y}}\|} \left(\|\boldsymbol{x}_{i}\|^2  +  \sum_{j\in\{1,\dots,m\}, \ j\neq i} \langle \boldsymbol{x}_{i},\boldsymbol{x}_{j} \rangle\right), \ i=1,\dots,m.
\]
According to assumption (\ref{eq:k-tuples:assumption}), the following holds
\[
\left\langle \frac{\bar{\boldsymbol{y}}}{\|\bar{\boldsymbol{y}}\|}, \boldsymbol{x}_{i} \right\rangle \geq \frac{1}{m \|\bar{\boldsymbol{y}}\|} \left((1-\varepsilon)^2 + \beta_2 (m-1) \right),
\]
and, respectively,
\[
\frac{1}{m}\left(1+(m-1)\beta_1 \right)\geq \langle\bar{\boldsymbol{y}},\bar{\boldsymbol{y}}\rangle = \|\bar{\boldsymbol{y}}\|^2 \geq \frac{1}{m}\left((1-\varepsilon)^2 + \beta_2 (m-1)\right)
\]
Let $(1-\varepsilon)^2  + \beta_2(m-1) > 0$ and $1 + \beta_1(m-1) > 0$. It is clear that for $\ell$, as defined by (\ref{eq:k-tuple:functional}), the following holds  for all $i=1,\dots,m$: $\ell(\boldsymbol{x}_{i})\geq 0$. $\square$

\bibliographystyle{plain}
\bibliography{HighDimensionalBrain}

\end{document}